\documentclass[journal,onecolumn,draftclsnofoot]{IEEEtran}

\ifCLASSINFOpdf
\usepackage[pdftex]{graphicx}
\else
\fi    

\usepackage{balance}  
\usepackage{cite}
\usepackage{graphicx}
\usepackage{float}
\usepackage{amsmath} 
\usepackage[utf8]{inputenc}
\usepackage[final]{pdfpages}  
\usepackage{pdfpages}
\usepackage{lipsum}
\usepackage{textcase}
\usepackage{url}
\usepackage{amsmath,esint}
\usepackage{epstopdf}
\usepackage{array} 
\usepackage{hyperref}
\usepackage{multirow}
\usepackage{subcaption}
\usepackage{booktabs}
\usepackage{amssymb}
\usepackage{makecell}

	\newcommand\blfootnote[1]{%
		\begingroup
		\renewcommand\thefootnote{}\footnote{#1}%
		\addtocounter{footnote}{-1}%
		\endgroup
	}
\newcolumntype{P}[1]{>{\centering\arraybackslash}p{#1}}
\newcolumntype{M}[1]{>{\centering\arraybackslash}m{#1}}

\begin{document}   
	
\title{Ultra-Wideband Air-to-Ground Propagation Channel Characterization in an Open Area}

\author{\IEEEauthorblockN{Wahab Khawaja\IEEEauthorrefmark{1},~Ozgur Ozdemir\IEEEauthorrefmark{1},~Fatih Erden\IEEEauthorrefmark{1},~Ismail Guvenc\IEEEauthorrefmark{1},~David W. Matolak\IEEEauthorrefmark{2}}

\IEEEauthorblockA{\IEEEauthorrefmark{1}Department of Electrical and Computer Engineering, North Carolina State University, Raleigh, NC}

\IEEEauthorblockA{\IEEEauthorrefmark{2}Department of Electrical Engineering, University of South Carolina, Columbia, SC}

Email: \{wkhawaj, oozdemi, ferden, iguvenc\}@ncsu.edu, matolak@cec.sc.edu,}

\maketitle
\blfootnote{This  work  has  been  supported  in  part  by  NASA  under  the  Federal Award ID number NNX17AJ94A and by National Science Foundation~(NSF) under the grant number CNS-1453678. Wahab Khawaja has also been supported via a Fulbright scholarship. We also thank Jianlin Chen from NCSU, for his help in the air-to-ground propagation measurements. This work was presented in part in the IEEE Aerospace Conference, Big Sky, MT, USA, March 2019~\cite{aerospace_wahab}.}

\begin{abstract}
 This paper studies the air-to-ground~(AG) ultra-wideband~(UWB) propagation channel through measurements between $3.1$~GHz to $4.8$~GHz using unmanned-aerial-vehicles~(UAVs). Different line-of-sight~(LOS) and obstructed-LOS scenarios and two antenna orientations were used in the experiments. Channel statistics for different propagation scenarios were obtained, and the Saleh-Valenzuela~(SV) model was found to provide a good fit for the statistical channel model. An analytical path loss model based on antenna gains in the elevation plane is provided for unobstructed UAV hovering and moving~(in a circular path) propagation scenarios.  
 

\end{abstract}

\IEEEpeerreviewmaketitle

\section{Introduction}
The use of civilian unmanned-aerial-vehicles~(UAVs) for applications such as video recording, surveillance, search and rescue, and hot spot communications has seen a surge in recent years. Compared to other aerial platforms, highly mobile UAVs have several advantages, including ease in take-off/landing and operability, multiple varieties of simple flight controls, small size, and affordable prices. These features make them excellent candidates for numerous current and future applications. One  promising applications is in the field of wireless communications, e.g., providing on-demand access to hot spot or disaster-hit areas~\cite{wahab_survey,merwaday2016improved}. A recent example was seen in Puerto Rico, after Hurricane Maria, where a large portion of the cellular infrastructure was damaged. UAVs were used there by AT\&T as base stations to provide cellular coverage~\cite{COW}.   

\begin{figure}[!t]
	\centering
	\vspace{-.2cm}
	\includegraphics[width=0.65\columnwidth]{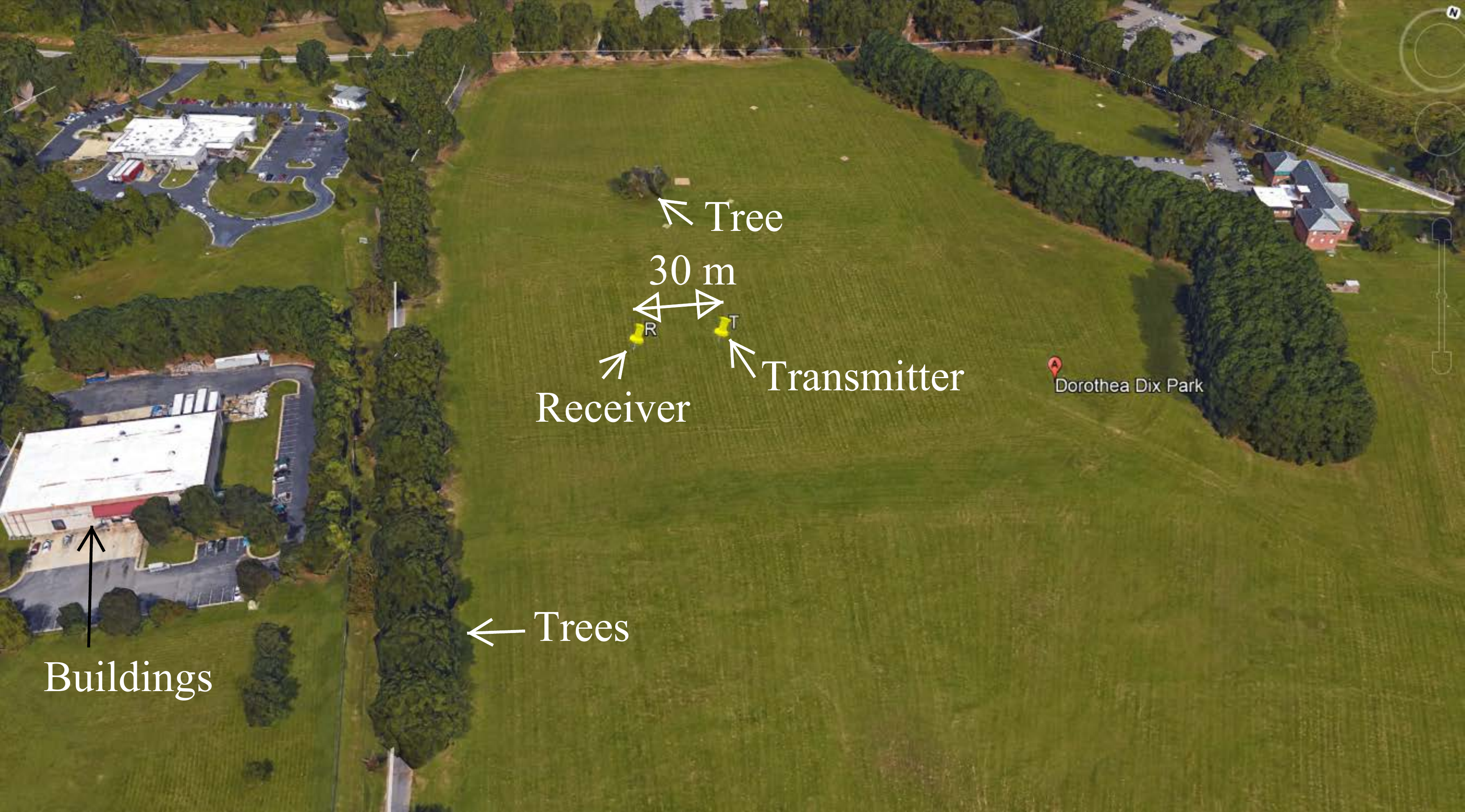}
	\vspace{-0.3cm}
	\caption{Channel measurement area from Google Earth.}\label{Fig:Google_maps}
\end{figure}

There are limited studies available on air-to-ground~(AG) propagation channel characterization in the literature~\cite{wahab_survey,Survey2}; here we cite a few examples. Narrowband AG propagation channel measurements using UAVs in an urban environment~\cite{Narrow_band1} consider a Loo model~(Rice and lognormal) for signal amplitude variations. A two-ray path loss model in an urban environment was observed to fit narrowband measurements carried out in an urban environment in~\cite{Narrow_band2}. Wideband AG propagation channel measurement campaigns in the L-band and C-band were performed for over water, mountains and hilly area, suburban, and near-urban environments in~\cite{Matolak_over_water,Matolak_mountains_hilly,Matolak_Suburban_urban}. Large scale and small scale propagation channel statistics and quasi-deterministic channel models were provided. Due to the difficulty of flying UAVs over populated and high-rise building areas, ray tracing simulations provide an alternative approach for channel characterization in these settings~\cite{wahab_GSMM,wahab_vtc,ray_tracing_urban}. Even with the literature in this area growing, there are limited AG propagation studies with UAVs in comparison to terrestrial, especially studies that specifically focus on the antenna radiation pattern effects~\cite{jianlin, antenna1, antenna2, wahab_GSMM}. 

To the best of our knowledge, there are also very few ultra-wideband~(UWB) AG propagation channel measurements in the literature, except for our previous studies~\cite{aerospace_wahab,wahab_uwb,jianlin}. In these studies, UWB AG propagation channel measurements and analyses were performed in different propagation environments. This present paper is a major extension of our previous work in~\cite{aerospace_wahab}. Here, our new contributions include a new analytical path loss model for the unobstructed UAV hovering and moving~(in a circular path) propagation scenarios. Ray tracing simulations were also conducted for the unobstructed UAV hovering scenario. In addition, analysis of best fits for the power delay profiles (PDPs) based on the Saleh-Valenzuela and single exponential models are provided. We also provide additional channel statistics: the root mean square delay spread~(RMS-DS) and Ricean $K$-factor. 

In~\cite{wahab_uwb}, UWB AG propagation channel characteristics using UAVs were studied in an open area and suburban area for different propagation scenarios. The empirical data was used to develop a statistical channel model. In~\cite{jianlin}, the effect of elevation angle for the received power at different UAV altitudes and antenna orientations were discussed for UWB propagation in an open area. The gain of the omni-directional dipole antenna in the elevation plane was modeled as a trigonometric function of elevation angle between the UAV and the ground station~(GS). 
Overall, the large bandwidth of UWB radio signals allows high temporal resolution of multipath components~(MPCs) that can provide detailed impulse response information for a given propagation environment. Studying these MPC characteristics can help in understanding the AG propagation channel for future broadband communications~\cite{wahab_survey}.

In this paper, we report on comprehensive channel measurements in an open area for three conditions: (1) unobstructed line-of-sight~(LOS) path~(no foliage) when the UAV is hovering; (2) obstructed line-of-sight~(OLOS) path due to foliage within the link while the UAV is hovering; and, (3) unobstructed LOS path while the UAV is moving in a circular trajectory. A snapshot of the measurement area from Google Maps is shown in Fig.~\ref{Fig:Google_maps}. The measurements were conducted using different horizontal and vertical distances of the transmitter~(TX) on the UAV to the receivers~(RXs) on the ground. Two different antenna orientations, vertical and horizontal~(also corresponding to the linear polarization), were used at the TX, whereas the orientation of the RX antennas was always vertical. The channel measurements were obtained using Time Domain P440 UWB radios operating in the frequency range $3.1$~GHz~-~$4.8$~GHz. All antennas are omni-directional in azimuth.

The main contributions of this AG measurement study can be summarized as follows:

\begin{itemize}
    \item The received power for the co-polarized antennas is mainly dependent on the antenna gain of the LOS component in the elevation plane for unobstructed UAV hovering scenario. For this scenario, we provide an analytical path loss model based on the antenna gain in the elevation plane, and we compare path loss measurements for this scenario with ray tracing simulation results.
    \item Antenna orientation mismatch results in higher path loss and RMS-DS, a larger number of weak multipath components (MPCs), and smaller Ricean $K$-factor than the co-polarized case. Moreover, the OLOS scenario introduces additional attenuation and MPCs due to foliage, resulting in further reduction in the $K$-factor.
    \item The motion of the UAV in an unobstructed circular path provides mitigation against antenna polarization mismatch effects in comparison to the unobstructed UAV hovering scenario. These include smaller path loss and RMS-DS for the unobstructed UAV moving scenario compared to the unobstructed UAV hovering scenario for the cross-polarized case.
    \item The Saleh Valenzuela~(SV) model is found to provide a better fit for the power delay profile~(PDP) than the single exponential model.
\end{itemize}

\begin{figure}[!t]
	\begin{subfigure}{0.47\textwidth}
	\centering
	\includegraphics[width=\textwidth]{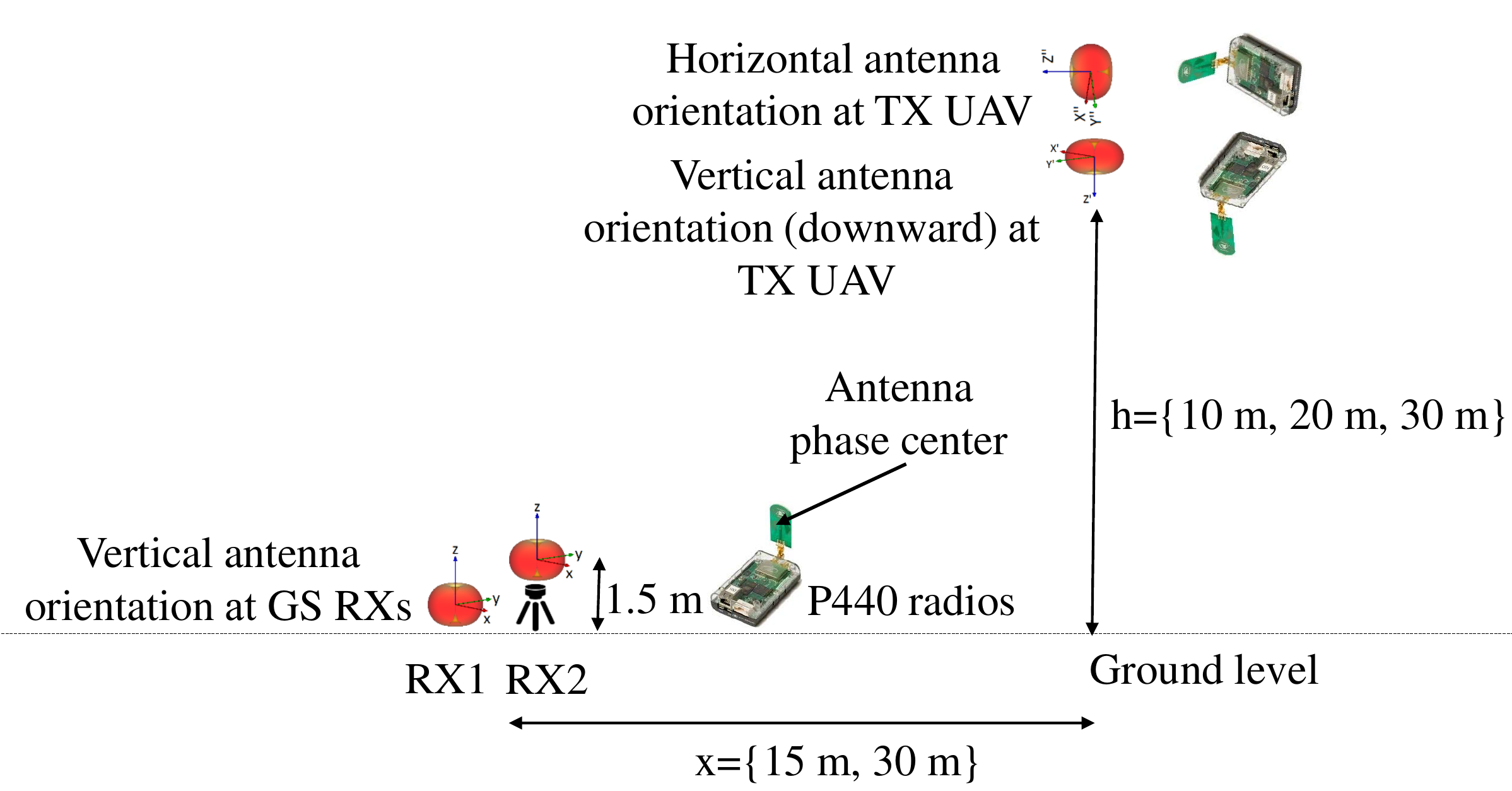} 
	\caption{} 
    \end{subfigure}			
	~\begin{subfigure}{0.47\textwidth}
	\centering
    \includegraphics[width=\textwidth]{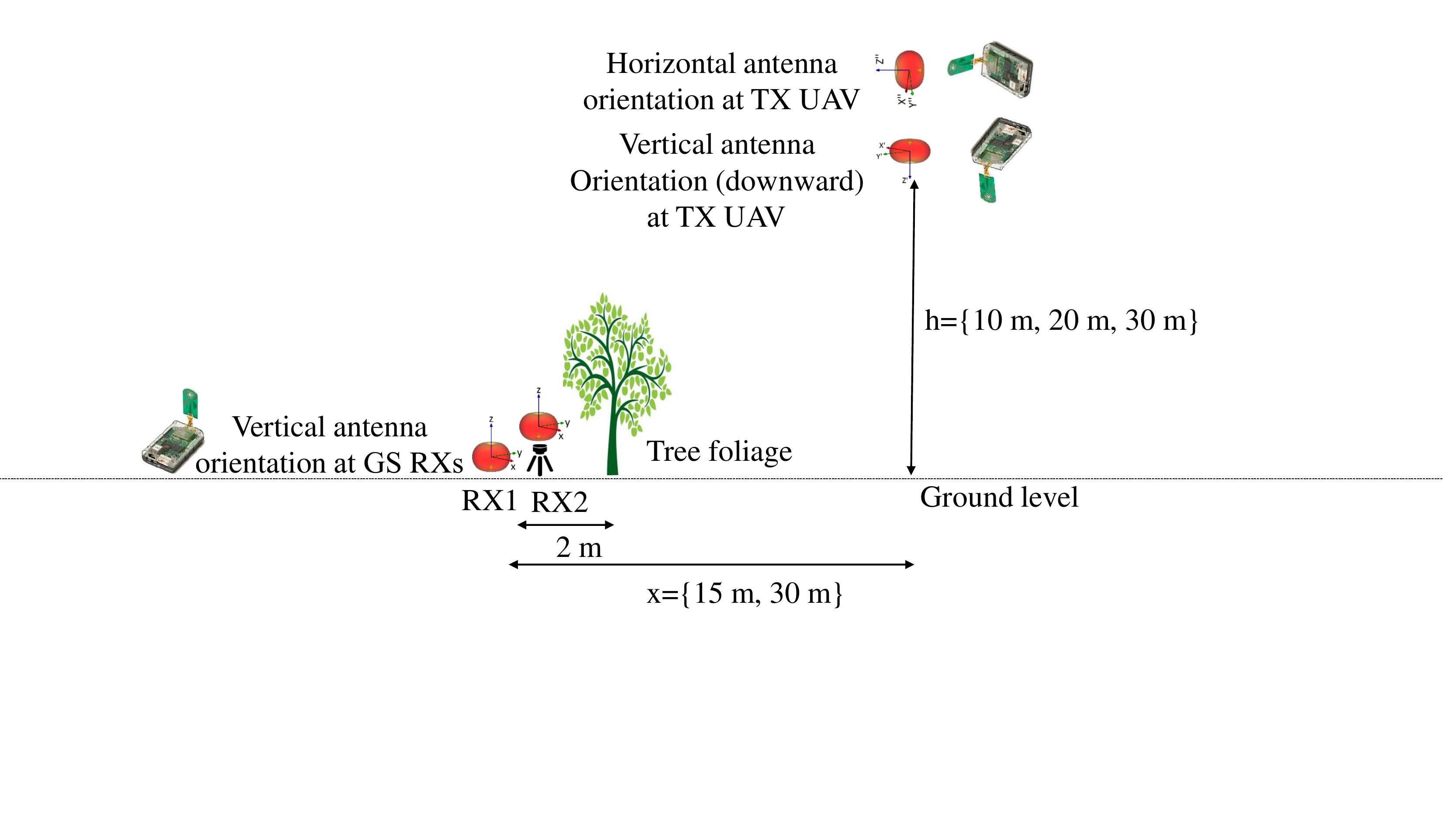}
	 \caption{}
     \end{subfigure}
	\begin{center}
     ~\begin{subfigure}{.47\textwidth}

    \includegraphics[width=\textwidth]{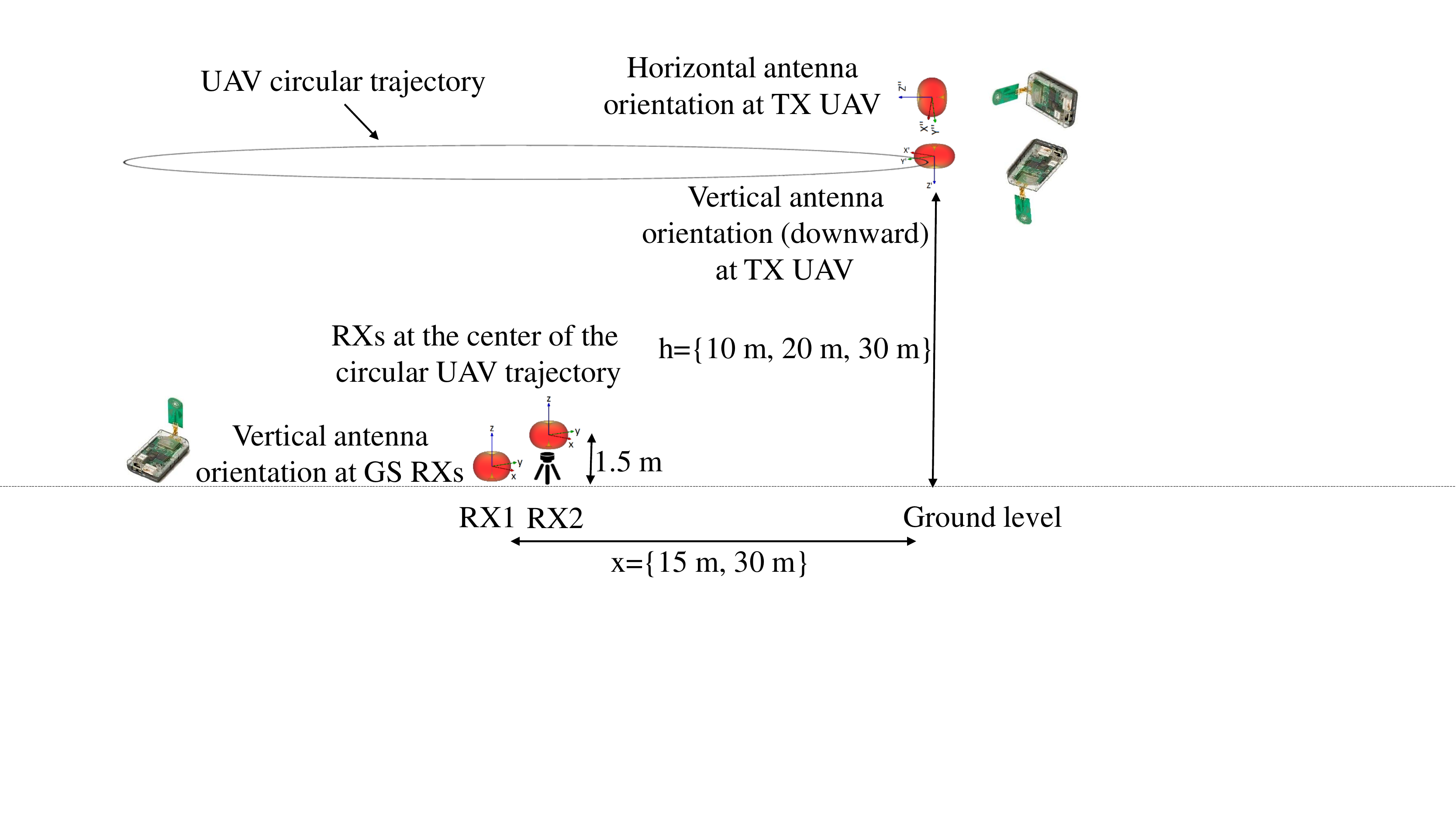}\caption{}
     \end{subfigure}\end{center}
     \caption{AG propagation scenarios in an open area for (a) unobstructed UAV hovering, (b) UAV hovering with tree foliage obstruction, (c) unobstructed UAV moving in a circular trajectory. Radiation patterns of the UWB antennas are also shown for different antenna orientations. }\label{Fig:Scenario_AG}
\end{figure}

The organization of this paper is as follows. Section~\ref{Section:Ch_measurements} explains the channel measurement setup and experiment scenarios. Using the data obtained from these experiments,  Section~\ref{Section:Ch_analysis_model} describes channel impulse response~(CIR) characterization, PDP model fitting, and small scale channel statistics. In Section~\ref{Section: Path_Loss_model}, antenna radiation pattern modeling, received power and path loss modeling are provided. This includes empirical path loss results as well as results from ray tracing simulations. Finally, Section~\ref{Section:Conclusions} concludes the paper.

\section{Channel Measurement Setup and Experiment Scenarios} \label{Section:Ch_measurements}
In this section, we describe the channel measurements conducted using Time Domain P440 radios and a DJI Phantom~4 UAV. The measurements were carried out in an open field close to North Carolina State University's Centennial campus. A Google map image of the measurement area is shown in Fig.~\ref{Fig:Google_maps}. The three different measurement scenarios are shown in Fig.~\ref{Fig:Scenario_AG}.  

\subsection{Channel Sounding with Time Domain P440 UWB Radios}
\begin{figure}[!t]
	\centering
	\includegraphics[width=0.65\columnwidth]{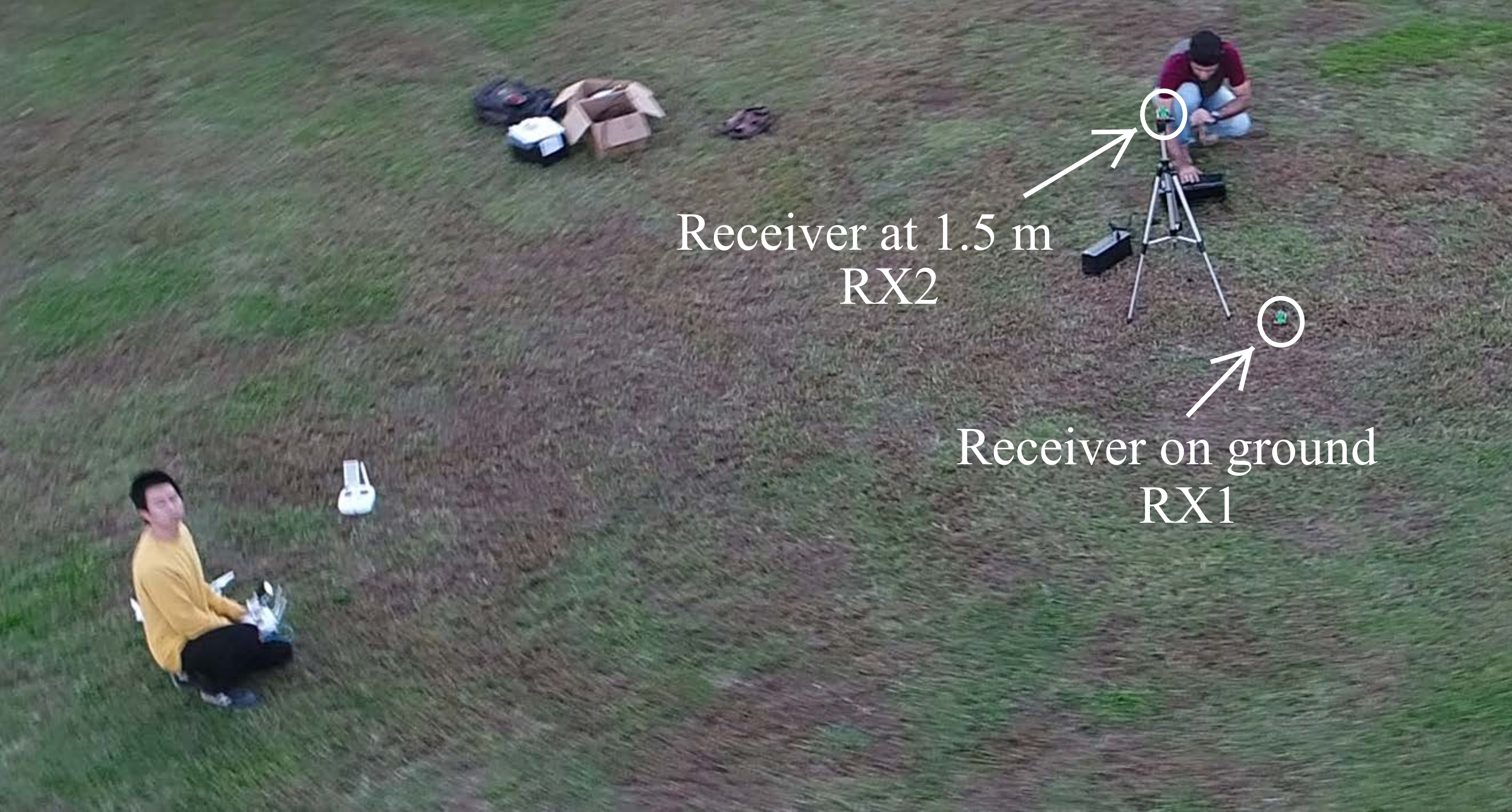}
	\caption{Channel measurements using DJI Phantom UAV and UWB P440 radios at two RX positions (snapshot from the UAV).}\label{Fig:Scenario_original}
\end{figure}

Channel sounding equipment is generally very bulky and often requires wired synchronization. This puts a constraint on the AG propagation channel measurements with conventional channel sounders using UAVs. Therefore, we used Time Domain P440 radios for UWB channel sounding since they provide easy to set up bi-static channel measurements. Additionally, no physical connection is required for synchronizing the TX and the RX. A central synchronizing clock signal is sent from the TX to the RX through packets. A very narrow pulse with approximately a Gaussian shape in the time domain is used. The duration of each pulse is $1$~ns and the repetition interval of the pulse is $100$~MHz, resulting in a scan duration of $100$~ns. The pulses are integrated into custom sized packets. The UAV used for the measurements was a DJI Phantom~4. Using the GS auto-pilot application\cite{DJI_GS}, the UAV flew exactly at the designated flight coordinates. A snapshot of the measurement environment is shown in Fig.~\ref{Fig:Scenario_original}.  

\begin{table}[!t]
	\begin{center}
		\caption{Specifications of channel sounding measurements.}\label{Table:Sounding_parameters}
        \begin{tabular}{@{} |P{4cm}|P{3cm}| @{}}
			\hline
			\textbf{Parameter}&\textbf{Parameter value}\\			
			\hline
			Operating frequency band& $3.1$~GHz~-~$4.8$~GHz \\
            \hline
            Center frequency& $3.95$~GHz \\
            \hline
            Pulse duration& $1$~ns \\
            \hline
            Dynamic range& $48$~dB \\
            \hline
            Pulse repetition rate& $10$~MHz \\
            \hline
            Noise figure at RX& $4.8$~dB \\
            \hline
            RX sensitivity& $-104$~dBm \\
            \hline
            RX time bin resolution& $1.9073$~ps \\
            \hline
            RX waveform measurements at& $32$ time bin interval \\
            \hline
            Communication link&Packet communication \\
            \hline 
            Antenna type at TX and RX& Planar elliptical dipole \\
            \hline
            Polarization& Vertical \\
            \hline
            Antenna pattern& Omni-directional in azimuth plane~($\pm 1.5$~dB) \\
            \hline
            Voltage standing wave ratio& $1.75:1$ \\
            \hline
            Antenna phase response& Linear \\
            \hline
            
	\end{tabular}
		\end{center}
\end{table}

Due to the coherent operation of TX and RX, the signal to noise ratio~(SNR) can be adjusted by changing the integration duration, i.e., the number of pulses  per packet. By increasing the pulse integration period, we can achieve longer ranges due to higher SNR. This can help in overcoming the power emission limitations by FCC\cite{FCC_emission}. However, a larger pulse integration period yields lower data rates, resulting in fewer channel scans captured in a given timing window. It also requires our channel to be time invariant for a longer period. In our experiment, we have used a pulse integration of $1024$ pulses per packet. This value ensures that we capture channel scans in a timing window without significant change of the propagation channel and at a reasonable link distance.

In addition to emission requirements by the FCC for UWB, there are two main factors affecting the SNR of the received signal. These factors limit the extraction of the CIR using the CLEAN algorithm~\cite{CLEAN}, which requires a given threshold of SNR. First is the preference to use omni-directional antennas instead of directional antennas for AG communications, and the second is the measurement noise. Omni-directional antennas with small antenna gain will be affected more by variations in the surrounding environment than will directional antennas. These variations may also be larger for aerial platforms than for terrestrial. Second, we observed high measurement noise from the equipment on-board the UAVs themselves in comparison to that observed at the GSs. This is mainly due to noise generated from the motors and propellers, vibrations on-board the UAV, and possibly other ambient effects, e.g., high temperatures experienced on-board the UAV, especially at higher UAV heights during a sunny day. These factors increase the RX noise, causing more frequent loss of transmitted packets, hence requiring larger coherent pulse aggregation per packet.

The UWB radios used in the experiment operate in the bi-static mode with a single transmit and receive antenna. In this mode, the TX continuously sends packets at an inter-packet delay of $10$~ms. A rake RX is used with a delay bin resolution of $1.9073$~ps. A standard $32$~bin duration is maintained between two measurements i.e., each measurement sample is processed after $61$~ps. The operating frequency range is $3.1$~GHz~-~$4.8$~GHz with an effective bandwidth of $1.7$~GHz~\cite{Time_Domain}.

\begin{figure}[!t]
	\begin{subfigure}{0.5\textwidth}
	\centering
	\includegraphics[width=\textwidth]{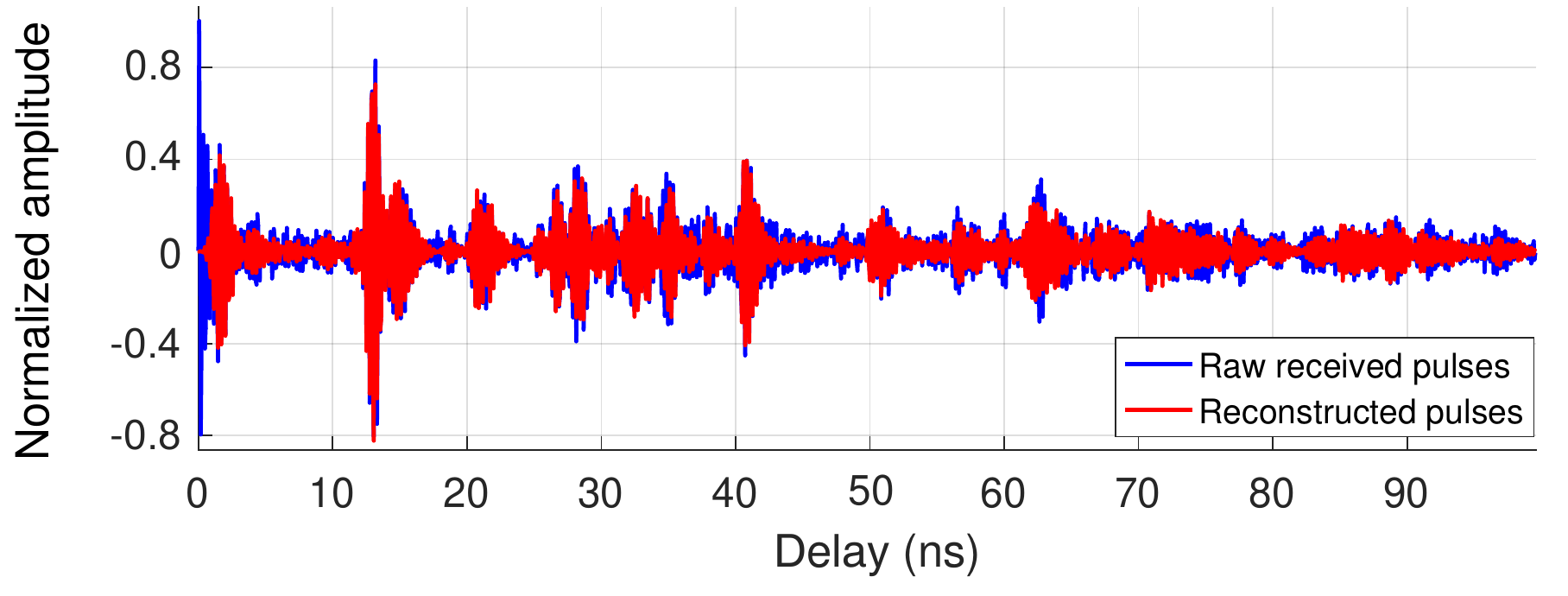} 
	\caption{}
    \end{subfigure}			
	\begin{subfigure}{0.5\textwidth}
	\centering
    \includegraphics[width=\textwidth]{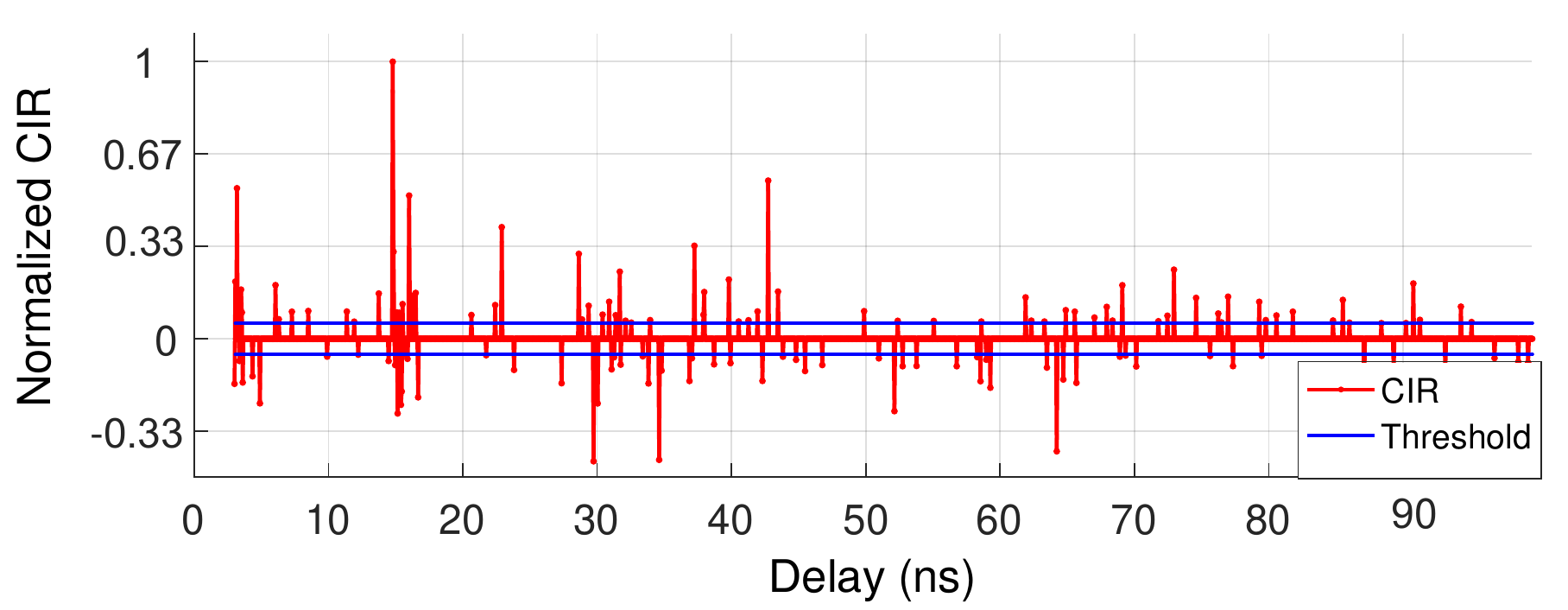}
	 \caption{}
     \end{subfigure}
     \caption{(a) Raw received and reconstructed pulses at the RX, (b) Channel impulse response with respect to delay obtained from raw received pulses at the RX.}\label{Fig:pulse_tx_rx}
\end{figure}

The received raw pulses are shown in Fig.~\ref{Fig:pulse_tx_rx}(a) in blue, whereas the reconstructed pulses shown in red are obtained by convolving the CIR shown in Fig.~\ref{Fig:pulse_tx_rx}(b) with the template waveform. The CIR in Fig.~\ref{Fig:pulse_tx_rx}(b) is obtained by deconvolving the received pulses with the template waveform. The blue horizontal lines indicate the amplitude threshold of the MPCs selected at $20\%$ of the peak amplitude. The channel sounding parameters are provided in Table~\ref{Table:Sounding_parameters}.

\subsection{Propagation Scenarios for Measurements}  
The experiments were designed to explore the UWB AG propagation channel characteristics in a typical open area. The three  propagation scenarios are illustrated in Fig.~\ref{Fig:Scenario_AG}. For the first scenario, there is no obstruction between the TX and the RX direct path while the UAV is hovering. For the second OLOS scenario, the TX and RXs are placed such that there is a medium-sized tree of height approximately $8$~m between them, as shown in Fig.~\ref{Fig:Scenario_AG}(b). The branches and leaves of the tree partially obstruct, scatter and diffract the transmitted energy. In the third scenario, measurements were taken while the UAV was moving in a circle at constant altitude, with RXs at the center. The velocity of the UAV was set at $6.1$~m/s and the TX orientation with respect to the UAV is kept constant~(i.e., we do not use any gimbal at the UAV to vary antenna direction). The motion in a circle ensures that distance remains constant between the TX and the RXs.

In all three propagation scenarios, two antenna orientations were used for the TX on the UAV~(vertical and horizontal, corresponding to the polarization). The antennas at the RXs were always vertically oriented as shown in Fig.~\ref{Fig:Scenario_AG}. For the first antenna orientation, the TX antenna was aligned vertically such that the antenna boresight~(with phase center in the middle) was facing the boresight of the RX antenna, when Tx and Rx were at the same height. This co-polarized antenna orientation at the RX and TX is called vertical-vertical and denoted VV. For the cross-polarized antenna orientation, the TX antenna was rotated $90^{\circ}$ for a vertical-horizontal~(VH) orientation. The VH antenna orientation was chosen in order to study the effects of the antenna orientation change on the channel characteristics for highly maneuverable UAVs. For both VV and VH antenna orientation, three UAV heights of $10$~m, $20$~m, and $30$~m at two horizontal distances of $x=15$~m and $x=30$~m were used. Two RXs, RX1 and RX2, were placed close to each other at heights of $10$~cm and $1.5$~m, respectively, from the ground.

\section{Modeling the Channel Impulse Response} \label{Section:Ch_analysis_model}

In this section, the channel impulse response (CIR) is analyzed. We develop model fits for PDPs and analyze multipath~(small scale) channel statistics, using the measurement data that has been collected as described in Section~\ref{Section:Ch_measurements}.

\subsection{Channel Impulse Response Model} \label{Section:CIR}

A generic model for the CIR of a time-varying wireless channel can be written as:
\begin{equation}
H(n) = \sum_{m = 0}^{M-1}\alpha_{m}(n) \exp\Big(j\varphi_{m}(n)\Big)\delta\Big(n-\tau_{m}(n)\Big), \label{Eq:CIR_general}
\end{equation}
where $M$ is the total number of MPCs, $\alpha_{m}(n)$, $\varphi_{m}(n)$ and $\tau_{m}(n)$ are the amplitude, phase and time of arrival~(TOA) or delay, respectively, of the $m^{\rm th}$ MPC at time instant $n$. For the UWB propagation channel, we observed clustering of the MPCs in the time domain due to resolvable reflections from individual scatterers. Therefore, considering $N_{\rm C}$ clusters and assuming the SV model~\cite{saleh}, the CIR expression in~(\ref{Eq:CIR_general}) can be modified to capture the clustering behavior as follows: An example CIR is provided in Fig.~\ref{Fig:pulse_tx_rx}, where we can observe clustered received power as a function of delay that supports the SV claim. 
\begin{equation}
H(n)=\sum_{l = 0}^{N_{\rm C}-1}\sum_{m=0}^{M_{l}-1}\alpha_{l,m}(n) \exp\Big(j\varphi_{l,m}(n)\Big)\delta\Big(n-T_l(n)-\tau_{l,m}(n)\Big), \label{Eq:Eq_CIR1}
\end{equation}
where $N_{\rm C}$ is the total number of clusters, $M_{l}$ is the total number of MPCs within the $l^{\rm th}$ cluster, $\alpha_{l,m}(n)$, $\varphi_{l,m}(n)$ and $\tau_{l,m}(n)$ are the amplitude, phase and arrival times, respectively, of the $m^{\rm th}$ MPC of the $l^{\rm th}$ cluster at time instant $n$, and $T_l(n)$ is the arrival time of the $l^{\rm th}$ cluster. For our specific measurements, the channel can be considered  time-invariant since all local scattering objects were motionless, and when the UAV is moving, the velocity is very small at $6.1$~m/s. Therefore, the CIR expression (\ref{Eq:Eq_CIR1}) can be simplified as follows:
\begin{equation}
H(n)=\sum_{l = 0}^{N_{\rm C}-1}\sum_{m=0}^{M_{l}-1}\alpha_{l,m} \exp\Big(j\varphi_{l,m}\Big)\delta\Big(n-T_l-\tau_{l,m}\Big). \label{Eq:Eq_CIR2}
\end{equation}

The mean square value of the $m^{\rm th}$ MPC of the $l^{\rm th}$ cluster is given in terms of the first MPC as:
\begin{equation}
\overline{\alpha_{l,m}^{2}} = \overline{|\alpha_{0,0}|^2}\exp({-T_l}{\eta})\exp({-\tau_{l,m}}{\gamma_l}), \label{Eq:Eq_Pwr} \end{equation}
 where $\overline{|\alpha_{0,0}|^2}$ is a mean power gain of the first path of the first cluster, and $\eta$ and $\gamma_l$ are the cluster and MPC power decay constants, respectively. Comparing (\ref{Eq:CIR_general}) and (\ref{Eq:Eq_CIR1}), the SV model converges to single exponential if $\gamma_{l}=0$.  

The arrival of the clusters and MPCs within each cluster can be modeled by Poisson processes~\cite{saleh,uwb_book_prof,report}, with respective arrival rates, $\chi$ and $\varsigma$ observed during the excess delay window. The inter-arrival times of clusters and MPCs are independent and can be fitted with an exponential distribution function~\cite{saleh} as:
\begin{align}
p(T_l|T_{l-1}) =& \chi \exp \big[-\chi(T_l - T_{l-1})\big], \label{Eq:Eq_Arrival_cluster}\\
p(\tau_m|\tau_{m-1}) =& \varsigma \exp \big[-\varsigma(\tau_m - \tau_{m-1}) \big].\label{Eq:Eq_Arrival_MPC}
\end{align}


\subsection{Analysis of Channel Impulse Response Measurements}
In this section, we consider the SV CIR model described in Section~\ref{Section:CIR} for characterizing our AG UWB channel measurements. The channel model parameters were obtained from the empirical results with the SV models characterized through (\ref{Eq:Eq_CIR2})--(\ref{Eq:Eq_Arrival_MPC})~\cite{uwb_book_prof}. The statistical propagation channel model parameters obtained from these equations are provided in Table~\ref{Table:Parameters_open}, \ref{Table:Parameters_foliage}, and \ref{Table:Parameters_moving} for the three scenarios in Fig.~\ref{Fig:Scenario_AG}. It can be observed that the cluster arrival rate $\chi$ captured in (\ref{Eq:Eq_Arrival_cluster}) is the highest for the unobstructed UAV hovering scenario, followed by foliage obstructed and unobstructed UAV moving scenarios. Similarly, $\chi$ is larger for the VV antenna orientation than for the VH antenna orientation for all three scenarios at both RXs.

\begin{table*}[!h]
  \centering
\begin{tabular}{|p{0.2cm}|p{0.2cm}|p{0.2cm}|p{0.2cm}|p{0.2cm}|p{0.2cm}|p{0.2cm}|p{0.2cm}|p{0.2cm}|}
\hline
		\multicolumn{1}{|c|}{}&\multicolumn{2}{|c|}{\textbf{RX1~(VV)}}&\multicolumn{2}{|c|}{\textbf{RX2~(VV)}}&\multicolumn{2}{|c|}{\textbf{RX1~(VH)}}&\multicolumn{2}{|c|}{\textbf{RX2~(VH)}} \\
		
			\hline
             \multicolumn{1}{|c|}{\textbf{Param.}}&\multicolumn{1}{|c|}{\textbf{\textit{x} = 15~m}}&\multicolumn{1}{|c|}{\textbf{\textit{x} = 30~m}}&\multicolumn{1}{|c|}{\textbf{\textit{x} = 15~m}}&\multicolumn{1}{|c|}{\textbf{\textit{x} = 30~m}}&\multicolumn{1}{|c|}{\textbf{\textit{x} = 15~m}}&\multicolumn{1}{|c|}{\textbf{\textit{x} = 30~m}}&\multicolumn{1}{|c|}{\textbf{\textit{x} = 15~m}}&\multicolumn{1}{|c|}{\textbf{\textit{x} = 30~m}}\\
\hline
\multicolumn{1}{|c|}{$N_{\rm C}$}& \multicolumn{1}{|c|}{$3.33$}& \multicolumn{1}{|c|}{$4$}& \multicolumn{1}{|c|}{$2.66$}& \multicolumn{1}{|c|}{$2$} & \multicolumn{1}{|c|}{$1.66$}& \multicolumn{1}{|c|}{$2.66$}& \multicolumn{1}{|c|}{$1.66$}& \multicolumn{1}{|c|}{$1.33$}\\
\hline
\multicolumn{1}{|c|}{$\chi~(\frac{1}{\rm ns})$}& \multicolumn{1}{|c|}{$.033$}& \multicolumn{1}{|c|}{$.04$}& \multicolumn{1}{|c|}{$.027$}& \multicolumn{1}{|c|}{$.02$} & \multicolumn{1}{|c|}{$.017$}& \multicolumn{1}{|c|}{$.027$}& \multicolumn{1}{|c|}{$.017$}& \multicolumn{1}{|c|}{$.013$}\\
\hline
\multicolumn{1}{|c|}{$\eta$}& \multicolumn{1}{|c|}{$.23$}& \multicolumn{1}{|c|}{$.186$}& \multicolumn{1}{|c|}{$.24$}& \multicolumn{1}{|c|}{$.16$}& \multicolumn{1}{|c|}{$.215$}& \multicolumn{1}{|c|}{$.16$}& \multicolumn{1}{|c|}{$.177$}& \multicolumn{1}{|c|}{$.171$}\\
\hline
\multicolumn{1}{|c|}{$\varsigma~(\frac{1}{ns})$}& \multicolumn{1}{|c|}{$.1$}& \multicolumn{1}{|c|}{$.06$}& \multicolumn{1}{|c|}{$.11$}& \multicolumn{1}{|c|}{$.06$}& \multicolumn{1}{|c|}{$.25$}& \multicolumn{1}{|c|}{$.15$}& \multicolumn{1}{|c|}{$.26$}& \multicolumn{1}{|c|}{$.2$}\\
\hline
\multicolumn{1}{|c|}{$\gamma$}& \multicolumn{1}{|c|}{$8.7$}& \multicolumn{1}{|c|}{$8.66$}& \multicolumn{1}{|c|}{$5.5$}& \multicolumn{1}{|c|}{$4.3$}& \multicolumn{1}{|c|}{$2.7$}& \multicolumn{1}{|c|}{$5.92$}& \multicolumn{1}{|c|}{$2.8$}& \multicolumn{1}{|c|}{$1.88$}\\
\hline
		\end{tabular}
\caption{UWB UAV channel model parameters averaged over UAV heights for an \textit{\textbf{open area}} while the UAV is  \textit{\textbf{hovering without obstruction}}.} \label{Table:Parameters_open}
\end{table*}

\begin{table*}[!h]
\centering
\begin{tabular}{|p{0.2cm}|p{0.2cm}|p{0.2cm}|p{0.2cm}|p{0.2cm}|p{0.2cm}|p{0.2cm}|p{0.2cm}|p{0.2cm}|}
\hline
			\multicolumn{1}{|c|}{}&\multicolumn{2}{|c|}{\textbf{RX1~(VV)}}&\multicolumn{2}{|c|}{\textbf{RX2~(VV)}}&\multicolumn{2}{|c|}{\textbf{RX1~(VH)}}&\multicolumn{2}{|c|}{\textbf{RX2~(VH)}} \\
			
			\hline
             \multicolumn{1}{|c|}{\textbf{Param.}}&\multicolumn{1}{|c|}{\textbf{\textit{x} = 15~m}}&\multicolumn{1}{|c|}{\textbf{\textit{x} = 30~m}}&\multicolumn{1}{|c|}{\textbf{\textit{x} = 15~m}}&\multicolumn{1}{|c|}{\textbf{\textit{x} = 30~m}}&\multicolumn{1}{|c|}{\textbf{\textit{x} = 15~m}}&\multicolumn{1}{|c|}{\textbf{\textit{x} = 30~m}}&\multicolumn{1}{|c|}{\textbf{\textit{x} = 15~m}}&\multicolumn{1}{|c|}{\textbf{\textit{x} = 30~m}}\\
\hline
\multicolumn{1}{|c|}{$N_{\rm C}$}& \multicolumn{1}{|c|}{$2$}& \multicolumn{1}{|c|}{$2$}& \multicolumn{1}{|c|}{$2$}& \multicolumn{1}{|c|}{$1.66$}& \multicolumn{1}{|c|}{$2$}& \multicolumn{1}{|c|}{$1.33$}& \multicolumn{1}{|c|}{$1.66$}& \multicolumn{1}{|c|}{$1.33$} \\
\hline
\multicolumn{1}{|c|}{$\chi~(\frac{1}{\rm ns})$}& \multicolumn{1}{|c|}{$.02$}& \multicolumn{1}{|c|}{$.02$}& \multicolumn{1}{|c|}{$.02$}& \multicolumn{1}{|c|}{$.017$}& \multicolumn{1}{|c|}{$.02$}& \multicolumn{1}{|c|}{$.013$}& \multicolumn{1}{|c|}{$.017$}& \multicolumn{1}{|c|}{$.013$} \\
\hline
\multicolumn{1}{|c|}{$\eta$}& \multicolumn{1}{|c|}{$.212$}& \multicolumn{1}{|c|}{$.21$}& \multicolumn{1}{|c|}{$.24$}& \multicolumn{1}{|c|}{$.23$}& \multicolumn{1}{|c|}{$.214$}& \multicolumn{1}{|c|}{$.16$}& \multicolumn{1}{|c|}{$.198$}& \multicolumn{1}{|c|}{.2} \\
\hline
\multicolumn{1}{|c|}{$\varsigma~(\frac{1}{ns})$}& \multicolumn{1}{|c|}{$.14$}& \multicolumn{1}{|c|}{$.175$}& \multicolumn{1}{|c|}{$.27$}& \multicolumn{1}{|c|}{$.21$}& \multicolumn{1}{|c|}{$.34$}& \multicolumn{1}{|c|}{$.34$}& \multicolumn{1}{|c|}{$.3$}& \multicolumn{1}{|c|}{$.34$} \\
\hline
\multicolumn{1}{|c|}{$\gamma$}& \multicolumn{1}{|c|}{$1.3$}& \multicolumn{1}{|c|}{$1.11$}& \multicolumn{1}{|c|}{$.985$}& \multicolumn{1}{|c|}{$1.34$}& \multicolumn{1}{|c|}{$.77$}& \multicolumn{1}{|c|}{$.811$}& \multicolumn{1}{|c|}{$1.4$}& \multicolumn{1}{|c|}{$.74$} \\
\hline
		\end{tabular}
\caption{UWB UAV channel model parameters averaged over UAV heights for \textit{\textbf{open area obstructed by foliage}} while the UAV is \textit{\textbf{hovering}}.}\label{Table:Parameters_foliage}
\end{table*}

\begin{table*}[!h]
\centering
\begin{tabular}{|p{0.2cm}|p{0.2cm}|p{0.2cm}|p{0.2cm}|p{0.2cm}|p{0.2cm}|p{0.2cm}|p{0.2cm}|p{0.2cm}|}
\hline
			\multicolumn{1}{|c|}{}&\multicolumn{2}{|c|}{\textbf{RX1~(VV)}}&\multicolumn{2}{|c|}{\textbf{RX2~(VV)}}&\multicolumn{2}{|c|}{\textbf{RX1~(VH)}}&\multicolumn{2}{|c|}{\textbf{RX2~(VH)}} \\
			\hline
             \multicolumn{1}{|c|}{\textbf{Param.}}&\multicolumn{1}{|c|}{\textbf{\textit{x} = 15~m}}&\multicolumn{1}{|c|}{\textbf{\textit{x} = 30~m}}&\multicolumn{1}{|c|}{\textbf{\textit{x} = 15~m}}&\multicolumn{1}{|c|}{\textbf{\textit{x} = 30~m}}&\multicolumn{1}{|c|}{\textbf{\textit{x} = 15~m}}&\multicolumn{1}{|c|}{\textbf{\textit{x} = 30~m}}&\multicolumn{1}{|c|}{\textbf{\textit{x} = 15~m}}&\multicolumn{1}{|c|}{\textbf{\textit{x} = 30~m}}\\
\hline
\multicolumn{1}{|c|}{$N_{\rm C}$}& \multicolumn{1}{|c|}{$2$}& \multicolumn{1}{|c|}{$1.66$}& \multicolumn{1}{|c|}{$1.66$}& \multicolumn{1}{|c|}{$1.33$}& \multicolumn{1}{|c|}{$2$}& \multicolumn{1}{|c|}{$1$}& \multicolumn{1}{|c|}{$1.66$}& \multicolumn{1}{|c|}{$1$}\\
\hline
\multicolumn{1}{|c|}{$\chi~(\frac{1}{\rm ns})$}& \multicolumn{1}{|c|}{$.02$}& \multicolumn{1}{|c|}{$.017$}& \multicolumn{1}{|c|}{$.017$}& \multicolumn{1}{|c|}{$.013$}& \multicolumn{1}{|c|}{$.02$}& \multicolumn{1}{|c|}{$.01$}& \multicolumn{1}{|c|}{$.017$}& \multicolumn{1}{|c|}{$.01$}\\
\hline
\multicolumn{1}{|c|}{$\eta$}& \multicolumn{1}{|c|}{$.14$}& \multicolumn{1}{|c|}{$.143$}& \multicolumn{1}{|c|}{$.2$}& \multicolumn{1}{|c|}{$.18$}& \multicolumn{1}{|c|}{$.15$}& \multicolumn{1}{|c|}{$.12$}& \multicolumn{1}{|c|}{$.205$}& \multicolumn{1}{|c|}{$.171$}\\
\hline
\multicolumn{1}{|c|}{$\varsigma~(\frac{1}{ns})$}& \multicolumn{1}{|c|}{$.1$}& \multicolumn{1}{|c|}{$.082$}& \multicolumn{1}{|c|}{$.084$}& \multicolumn{1}{|c|}{$.084$}& \multicolumn{1}{|c|}{$.14$}& \multicolumn{1}{|c|}{$.11$}& \multicolumn{1}{|c|}{$.16$}& \multicolumn{1}{|c|}{$.16$}\\
\hline
\multicolumn{1}{|c|}{$\gamma$}& \multicolumn{1}{|c|}{$1.87$}& \multicolumn{1}{|c|}{$1.87$}& \multicolumn{1}{|c|}{$3.6$}& \multicolumn{1}{|c|}{$5.2$}& \multicolumn{1}{|c|}{$1.76$}& \multicolumn{1}{|c|}{$2$}& \multicolumn{1}{|c|}{$2.04$}& \multicolumn{1}{|c|}{$1.31$}\\
\hline
		\end{tabular}
\caption{UWB UAV channel model parameters averaged over UAV heights for an \textit{\textbf{open area}} while the UAV is \textit{\textbf{moving}}.}\label{Table:Parameters_moving}
\end{table*}
\vspace{-0.35cm}
The arrival rate of MPCs $\varsigma$ in (\ref{Eq:Eq_Arrival_MPC}) is the largest for the foliage obstructed scenario. This is attributable to the multiple reflections from the tree~(branches, leaves, and trunk). Similarly, $\varsigma$ is larger for the VH than the VV antenna orientation at both RX1 and RX2. This is mainly due to a larger number of small powered MPCs for VH antenna orientation compared to VV antenna orientation.

The average of the cluster power decay constant $\eta$ from~(\ref{Eq:Eq_Pwr}) is approximately the same for all the propagation cases of the three scenarios. The value of $\eta$ is dependent on the peak power values of the individual clusters. For example, for the unobstructed UAV hovering scenario, we observe a larger number of peaks contributing to the overall $\eta$ due to a larger number of individual clusters. However, for the other two scenarios, we have a single or two clusters at maximum, where the  peak contribution is mainly from the first cluster. Moreover, we observe higher $\eta$ for the VV antenna orientation than for the VH antenna orientation at both RX1 and RX2. This is mainly due to faster power decay for the VV antenna orientation compared to the VH antenna orientation, where the power decay is relatively slower. In addition, for the unobstructed UAV moving scenario, we observe the smallest overall $\eta$, showing more uniform received power distribution compared to the other two scenarios.

Interesting observations can be made from the MPC power decay constant, $\gamma$ of (\ref{Eq:Eq_Pwr}), in three different scenarios. The value of $\gamma$ is the highest for the unobstructed UAV hovering scenario. The clusters observed in this case are of short duration with sharp power decay~(see Section~\ref{Section:Power_Clusters}). In the case of foliage, we observe only a few clusters. The power from the large duration foliage clusters decays slowly, resulting in overall smaller $\gamma$ . The value of $\gamma$ for the unobstructed UAV moving scenario is in between the other two scenarios. As with the cluster decay constant, $\gamma$ is larger for the VV antenna orientation than for the VH antenna orientation for both RXs as there is no dominant cross-polarized component.

\subsection{Saleh-Valenzuela versus Single Exponential Model PDP Fitting} \label{Section:SV_single}

\begin{figure}[!t]
	\begin{subfigure}{0.5\textwidth}
	\centering
	\includegraphics[width=\textwidth]{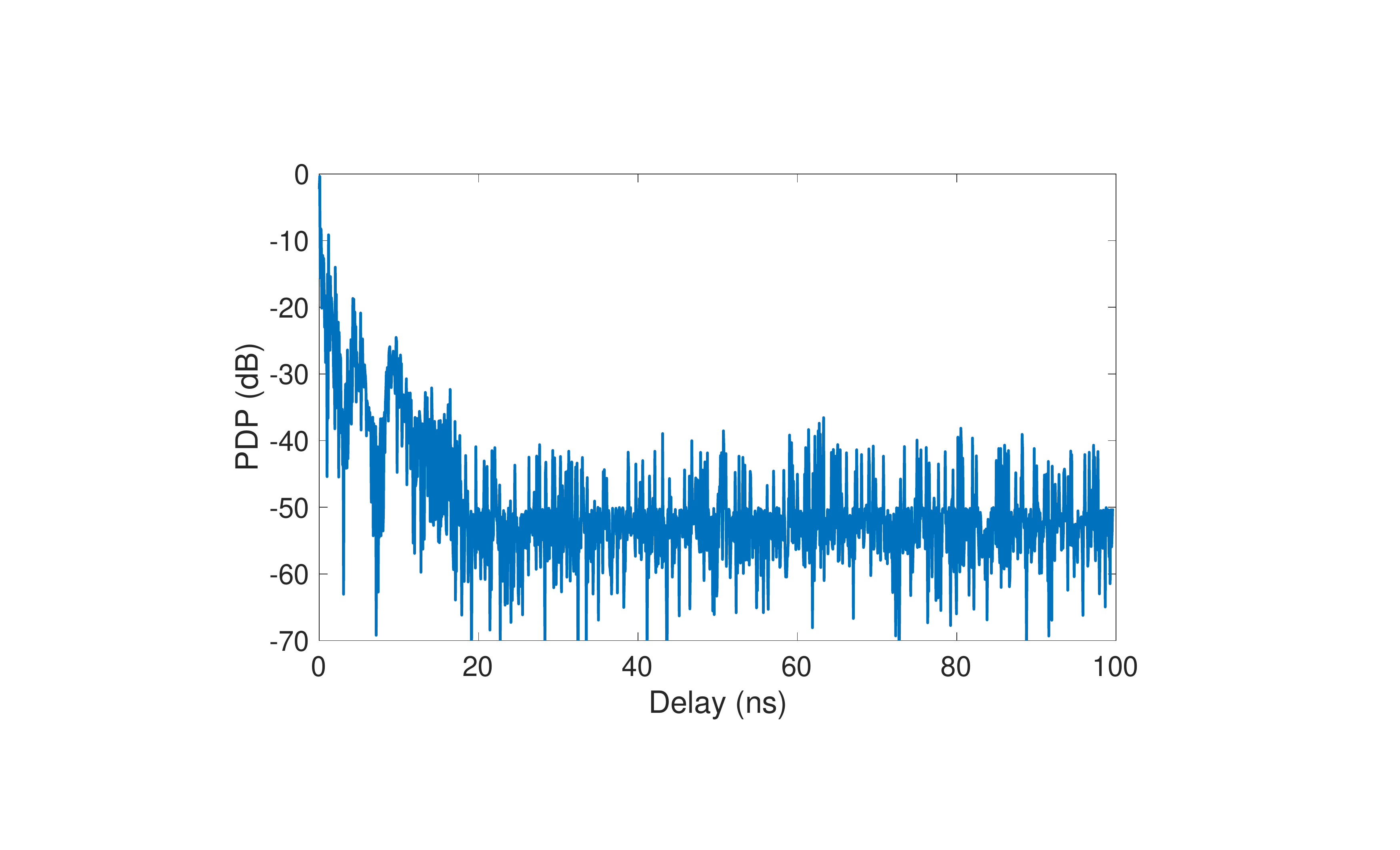} 
	\caption{}
    \end{subfigure}			
	\begin{subfigure}{0.5\textwidth}
	\centering
    \includegraphics[width=\textwidth]{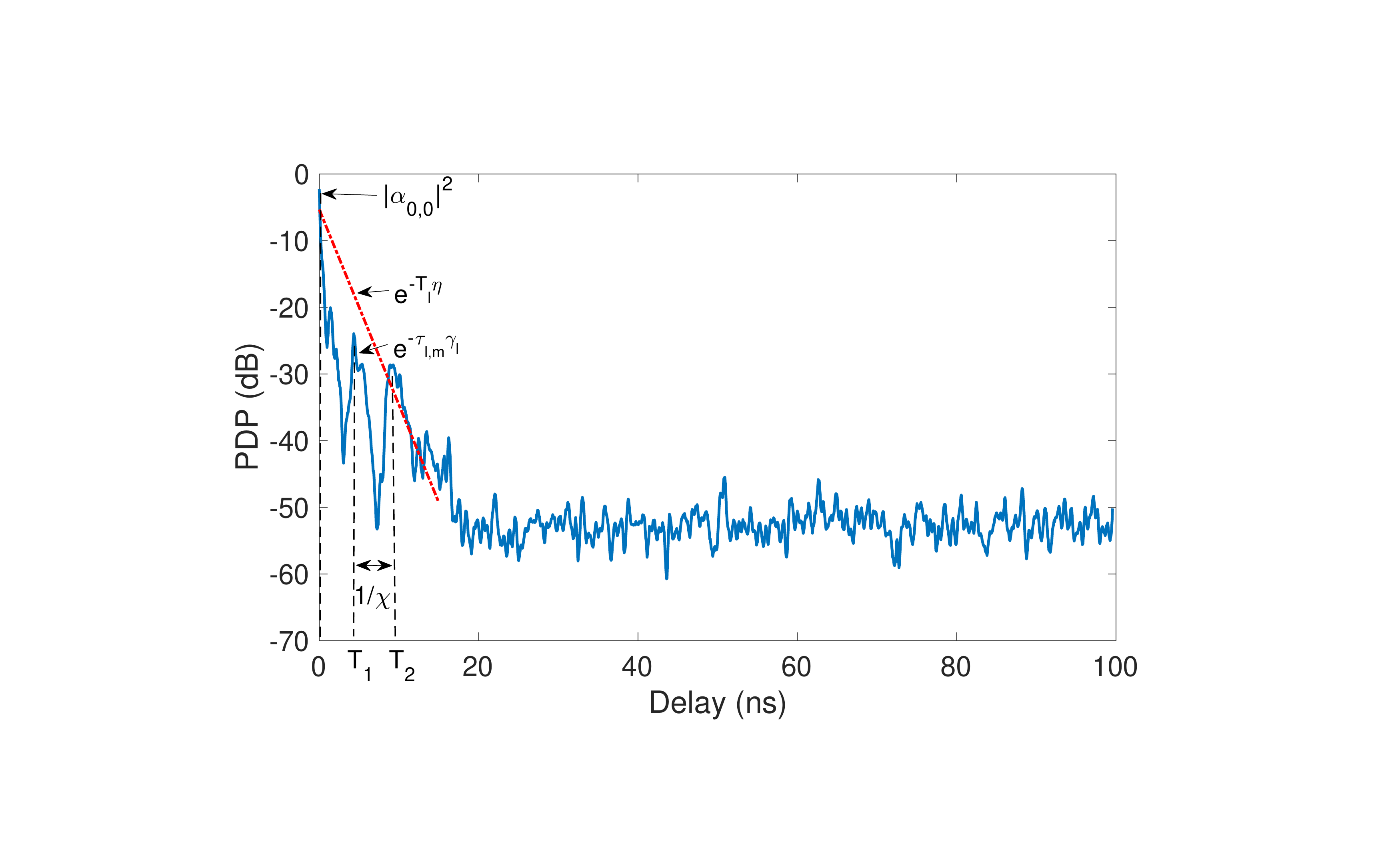}
	 \caption{}
     \end{subfigure}
     \caption{(a) Normalized and averaged empirical PDP, and (b) smoothed version. UAV is hovering at a height of $10$~m and horizontal distance $x = 30$~m from RX1 with VV antenna orientation. There is no foliage. }\label{Fig:PDP}
\end{figure}

The PDP is obtained from the CIR as $E(|H(n)|^2)$ from~\cite{two_ray2}, where $E$ is the expectation operator. A total of $50$ CIRs are collected for each scenario during a fixed time interval of $20$~s. In this subsection, we analyze the best fit for the empirical PDPs. The PDP for one of the measurement scenarios\footnote{For this case the UAV is hovering at a height of $10$~m. The horizontal distance between the TX and the RX is $x=30$~m and the antenna orientation is VV.} is shown in Fig.~\ref{Fig:PDP}. It can be observed that we have at least three distinct clusters. The cluster identification is performed by visually inspecting the PDP and make sure each cluster is at least $2.5$~ns long (for our 1 ns resolution) and there is an $8$~dB change from the peak of the cluster before the next cluster. These lower bounds ensure distinct classification of the clusters for our empirical data. The clusters detected using this approach are shown in Fig~\ref{Fig:PDP}(b). The $2.5$~ns duration and $8$~dB decay were chosen to enable identification of distinct clusters and avoid transient effects.

When the power is plotted in dB scale as in Fig.~\ref{Fig:PDP}, the exponential power decay in the absolute scale becomes a linear decay. Therefore, the PDP can be fit by a single linear fit~(logarithm of the exponential) function or by the SV model~\cite{saleh}, where individual linear fits are assigned for each cluster. Let $\tau_{l,i}$ and $P_{l,i}$~(in dB) represent the delay and power value obtained empirically at the $i^{\rm th}$ data point\footnote{Individual data points are not necessarily the MPCs. These data points must be above the given threshold set to be a valid MPC.} of the $l^{\rm th}$ cluster given as
\begin{align}
P_{l,i} = 10\log_{10}\Big(|\alpha_{0,0}|^2 \exp({-T_{l}}{\eta})\exp({-\tau_{l,i}}{\gamma_l})\Big)
\end{align}
for $i=0,1,2,\dots,N_{l}-1$, where $N_{l}$ is the total number of data points in the $l^{\rm th}$ cluster. Let $f^{(l)}(\tau_{l,i},\beta_{0},\beta_{1})$ represent the linear fit function for the $l^{\rm th}$ cluster given as
\begin{align}
f^{(l)}(\tau_{l,i},\beta_{0},\beta_{1}) = \beta_{0} + \beta_{1}\tau_{l,i}, 
\end{align}
where $\beta_{0}$ and $\beta_{1}$ are the y-intercept and slope of $f^{(l)}(\cdot)$, respectively. Then, the LS error $R_l$ for the $l^{\rm th}$ individual cluster can be written as follows:
\begin{align}
R_{l}(\beta_0,\beta_1)=\sum_{i=0}^{N_l-1}\bigg[f^{(l)}(\tau_{l,i},\beta_{0},\beta_{1})-P_{l,i} \bigg]^2,
\end{align}
which is the sum of the squared residuals between the linear fit $f^{(l)}(\tau_{l,i},\beta_{0},\beta_{1})$ and corresponding empirical data point values $P_{l,i}$ for the $l^{\rm th}$ cluster. Subsequently, the model parameters, $(\beta_0,\beta_1)$ can be extracted by jointly solving the following equations
$\frac{\partial R_{l}(\beta_0,\beta_1)}{\partial \beta_{0}}=0$ and $\frac{\partial R_{l}(\beta_0,\beta_1)}{\partial \beta_{1}}=0$.

\begin{figure*}
	\begin{subfigure}{0.5\textwidth}
	\centering
	\includegraphics[width=\textwidth]{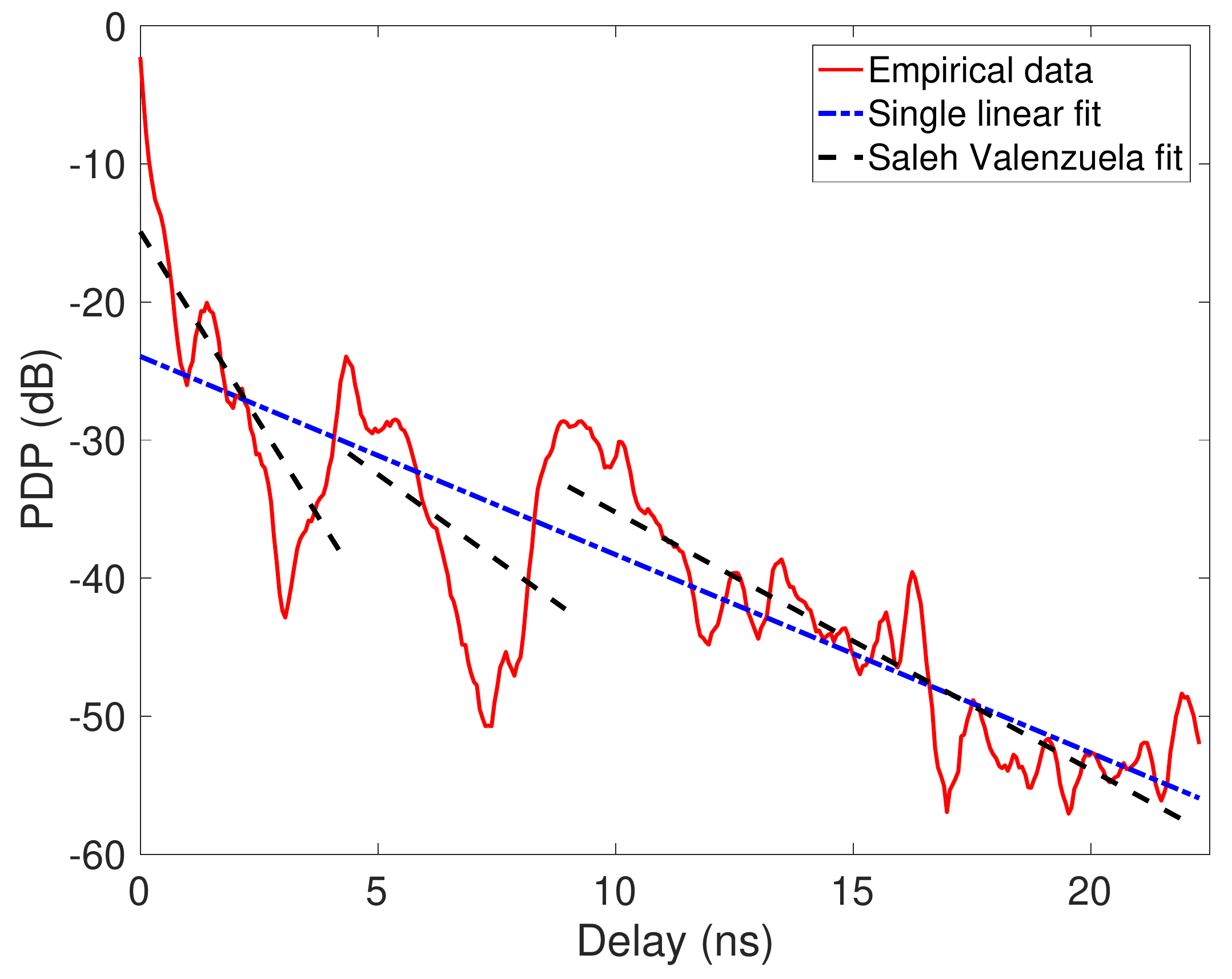}
	\caption{}
    \end{subfigure}			
	\begin{subfigure}{0.5\textwidth}
	\centering
    \includegraphics[width=\textwidth]{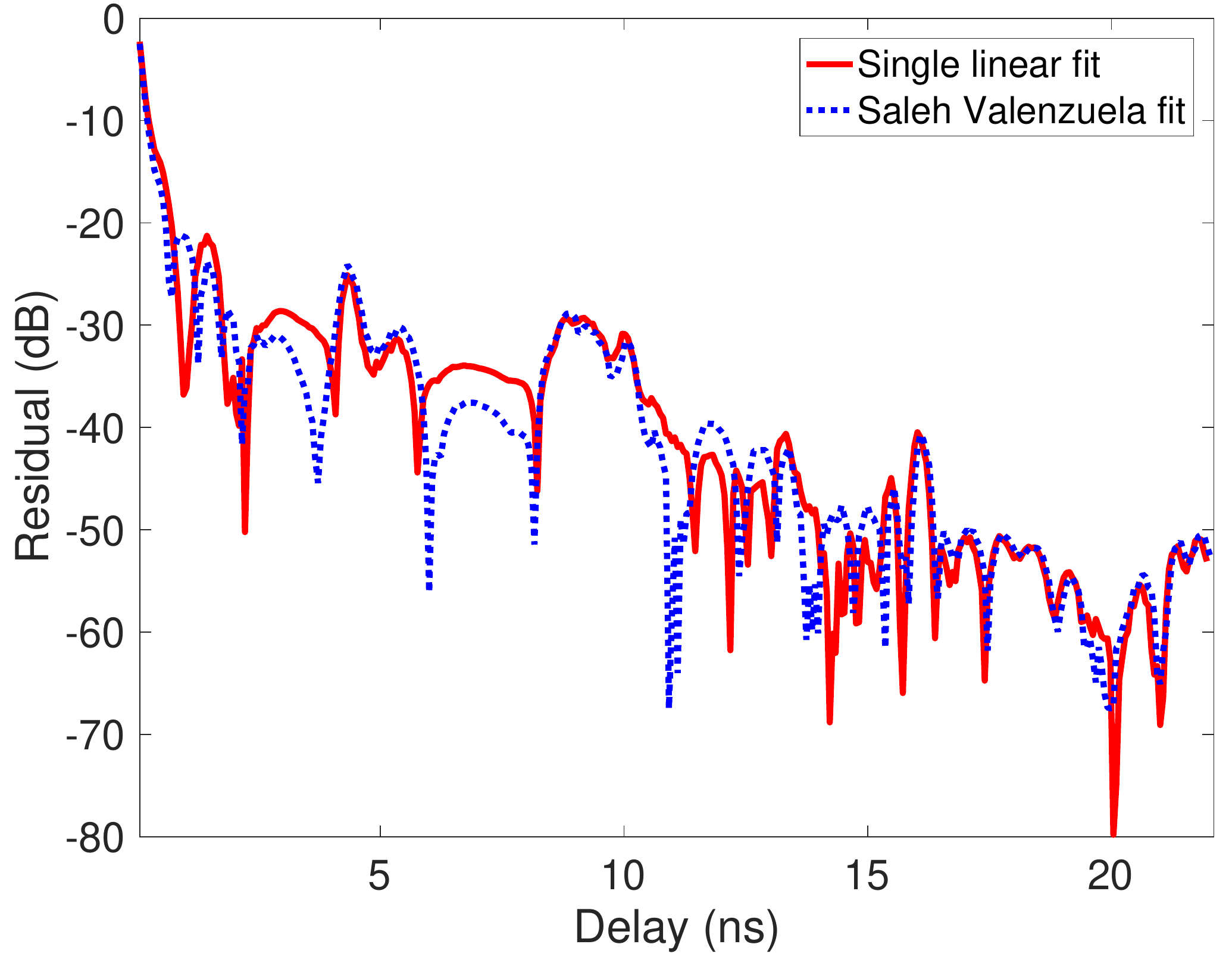}
	 \caption{}
     \end{subfigure}
       \caption{(a) Normalized and averaged empirical PDP, linear least square fitted with corresponding Saleh Valenzuela and single exponential. UAV is hovering at a height of $10$~m and horizontal distance $x = 30$~m from RX1. There is no foliage and antenna orientation is VV, (b) the absolute value of residuals for single linear and Saleh Valenzuela fit.} \label{Fig:Exp_SV_fit}
\end{figure*}

On the other hand, for the single cluster linear fitting, we have a single linear fit function for all the empirical data points in the PDP given as $f(\tau_{i},\beta_{0}',\beta_{1}') = \beta_{0}' + \beta_{1}'\tau_{i}$, where $\beta_{0}'$ and $\beta_{1}'$ are the y-intercept and slope of $f^{}(\cdot)$, respectively, for $i=0,1,2,...,N$, where $N$ is the total number of data points in the PDP. For this case, the empirical data is given as $P_i = |\alpha_{i}(\tau_i)|^2$ and the LS fit is obtained over the $N$ data points. The cluster-based individual linear fits~(based on the SV model) and the single linear fit to the empirical data for one of the propagation scenarios are shown in Fig.~\ref{Fig:Exp_SV_fit}(a). The absolute value of the residuals of the two fits is shown in Fig.~\ref{Fig:Exp_SV_fit}(b). The mean of the absolute residual for the single linear fit is $-23.58$~dB, whereas, for the SV model fit it is $-24.23$~dB. The smaller mean residual value for the SV model compared to the single linear fit suggests that the SV model provides a better fit to the empirical PDP than the single linear fit. Similar fittings can be obtained for other propagation scenarios.  

\subsection{Multipath Channel Analysis: Power Clusters} \label{Section:Power_Clusters}
A common phenomenon to observe during UWB propagation is the clustered reception of power~\cite{uwb_cluster}. In our outdoor open area environment with few obstacles and modest excess delay, we observe a small number of clusters in the PDP, as shown in Fig.~\ref{Fig:PDP}. The clusters are identified by visual inspection, using the distinct boundaries of power decay and rise discussed in Section~\ref{Section:SV_single}. The mean cluster count $N_{\rm C}$ captured in (\ref{Eq:Eq_CIR2}), is provided in Tables~\ref{Table:Parameters_open}, \ref{Table:Parameters_foliage}, and \ref{Table:Parameters_moving}. The unobstructed UAV hovering scenario has the largest mean cluster count followed by the foliage obstructed scenario and unobstructed UAV moving scenarios.

In the case of unobstructed UAV hovering without obstruction scenario shown in Fig.~\ref{Fig:Scenario_AG}(a), we observe essentially independent reflections from small scatterers near the RX, yielding several distinct clusters. However, in the case of UAV hovering with link obstructed by foliage as shown in Fig.~\ref{Fig:Scenario_AG}(b), usually only $2$ clusters are observed: one due to OLOS~(through the foliage) and the second mainly from the tree body~(diffraction around the tree crown and trunk). In addition, in case of foliage, the second cluster time delay is large compared to that in the other scenarios from the multiple reflections from foliage. In the unobstructed UAV moving scenario shown in Fig.~\ref{Fig:Scenario_AG}(c), we have a small number of clusters. The UAV moving scenario exhibits fewer clusters, likely due to spatial averaging.

 We observe larger mean cluster count for the VV antenna orientation than the VH antenna orientation for all propagation scenarios: the received power in the VH antenna orientation contains fewer strong reflected components, and this produces a more uniform distribution in delay than in the VV antenna orientation, hence the smaller VH cluster count. We also observe larger mean cluster count at RX1 than at RX2, likely due to reflections from the tripod body. The tripod body provides additional reflections and at the same time may help guide the energy towards the RX on the ground shown in Fig.~\ref{Fig:Scenario_original}.


\begin{figure}[!t]
	\centering
	\includegraphics[width=\textwidth]{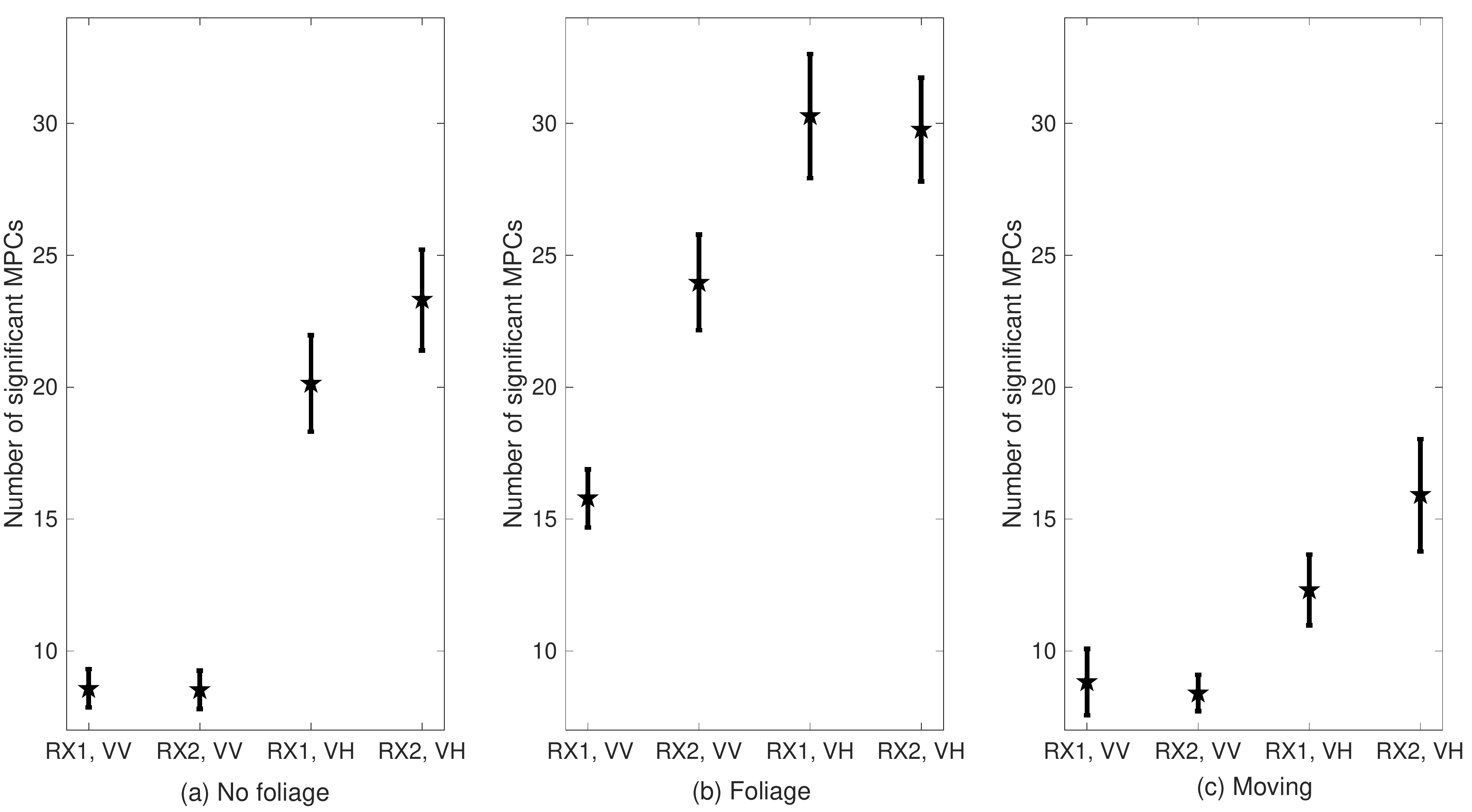}
	\caption{Average number of significant MPCs over multiple channel scans with 95\% confidence intervals for (a) unobstructed UAV hovering, (b) with foliage obstruction, and (c) unobstructed UAV moving on a circle. The average number of significant MPCs are obtained by averaging over UAV heights at respective horizontal distances for receiver positions RX1 and RX2 with VV and VH antenna orientations.}\label{Fig:Num_MPCs_mean}
\end{figure}

 \subsection{Multipath Channel Analysis: Number of Significant MPCs} \label{Section:Significant_MPCs}
We obtained the number of significant MPCs by retaining only the MPCs that were above the threshold of $20\%$ of the maximum MPC amplitude for a given CIR. These significant MPCs are counted for every scan of a given scenario. Then, they are averaged over the three UAV heights and two horizontal distances of $x = 15$~m and $x = 30$~m of a given scenario. The plot of the average number of significant MPCs with 95\% confidence intervals for the three flight conditions is shown in Fig.~\ref{Fig:Num_MPCs_mean}. It can be observed that we have a larger number of significant MPCs for VH antenna orientation than for the VV antenna orientation. This is because the low powered cross-polarized components are relatively larger when there is no LOS component  than the longer-delayed MPCs in the presence of a dominant co-polarized LOS component.

We observe the largest number of significant MPCs for the foliage obstructed scenario. On the other hand, we have the smallest number of significant MPCs for the unobstructed UAV moving and hovering scenarios with VV antenna orientation. This is because the presence of the dominant LOS component means that only a small number of large power MPCs exceed our threshold. We also observe a larger number of MPCs for RX2 than for RX1 because of RX2's better ground clearance. The potential scatterers near the RXs that provided the MPCs are the tripod, measuring equipment, two humans, and nearby chairs.  

\begin{figure}[!t]
	\centering
	\includegraphics[width=\textwidth]{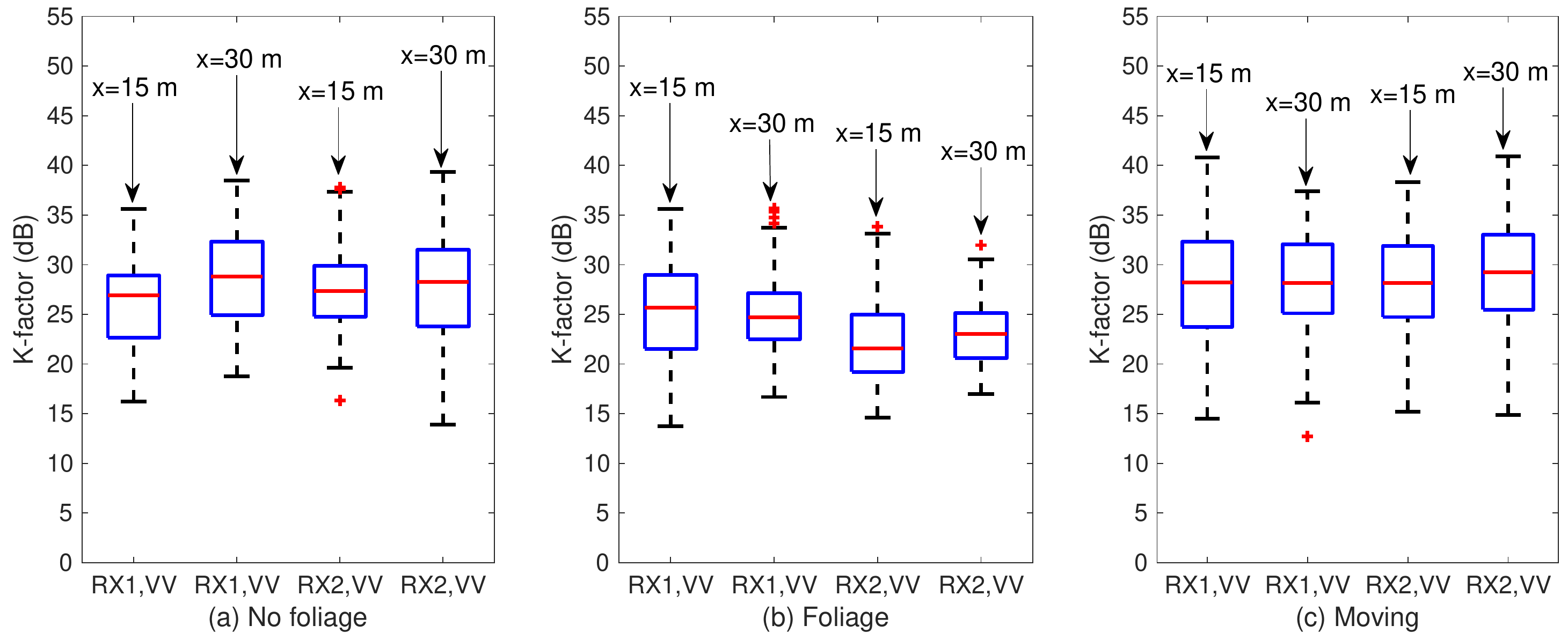}
	\caption{A box plot of Ricean $K$-factor for VV antenna orientation. Each box~(bounded between the $25^{\rm th}$ and $75^{\rm th}$ percentile and median in the center) represents the $K$-factor variation at all three UAV heights for (i) unobstructed UAV hovering, (ii) with foliage obstruction, and (iii) unobstructed UAV moving on a circle. The outliers are indicated by red $+$ sign.}\label{Fig:K_factor_mean}
\end{figure}

\subsection{Multipath Channel Analysis: $K$-factor}
The Ricean $K$-factor is obtained using the well-known equation, $K=10\log_{10}\frac{A^2}{2\sigma^2}$, where $A^2$ is the power of the LOS component. The LOS component is obtained as the first arriving component of the received waveform with highest power. The $\sigma^2$ term represents the power of the remaining MPCs, whose distribution is assumed Rayleigh. The Ricean $K$-factor at RX1 and RX2 for VV antenna orientations and different propagation scenarios is shown in Fig.~\ref{Fig:K_factor_mean}. Each box~(bounded between the $25^{\rm th}$ and $75^{\rm th}$ percentile with median in the center) represents the $K$-factor variation at all three UAV heights. The $K$-factor is not evaluated for VH antenna orientation due to weaker or absent LOS component in the channel scans. It can be observed that we have smaller mean $K$-factor for the foliage obstructed scenario than for the unobstructed scenarios, due to a weaker LOS component and larger number of MPCs for the foliage obstructed scenario. Moreover, for the UAV moving scenario, the $K$-factor has higher mean than the other two scenarios and the $K$-factor varies less than in the to other scenarios. 
 The larger value of $K$-factor for the moving scenario is likely attributable to some of the spatial averaging of the MPCs during the circular motion around the fixed receiver position. 
 
 \begin{figure}[!t]
	\centering
	\begin{subfigure}{0.75\textwidth}
	\centering
	\includegraphics[width=\columnwidth]{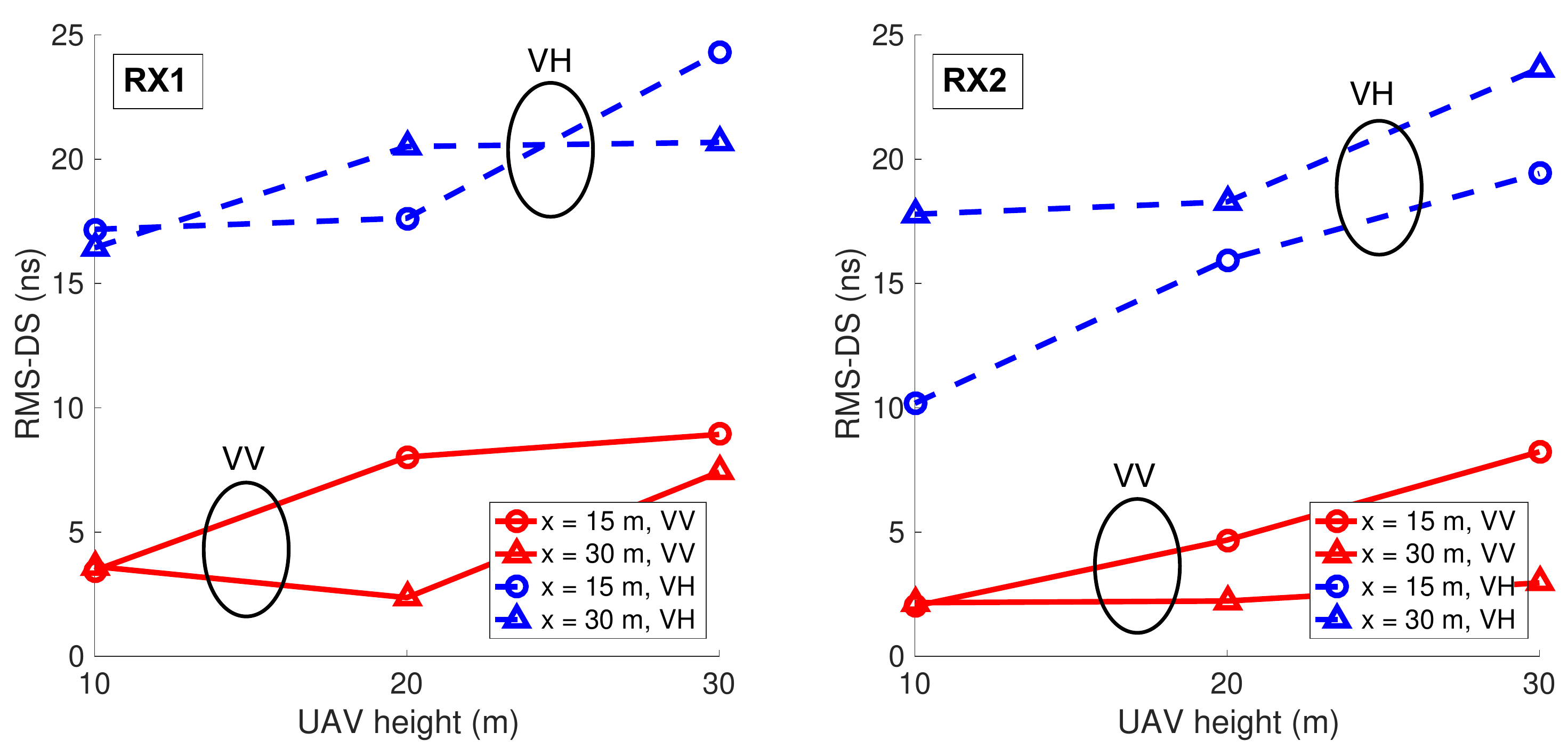} 
	\caption{Unobstructed UAV hovering.}
    \end{subfigure}			
	\begin{subfigure}{0.75\textwidth}
	\centering
	\includegraphics[width=\columnwidth]{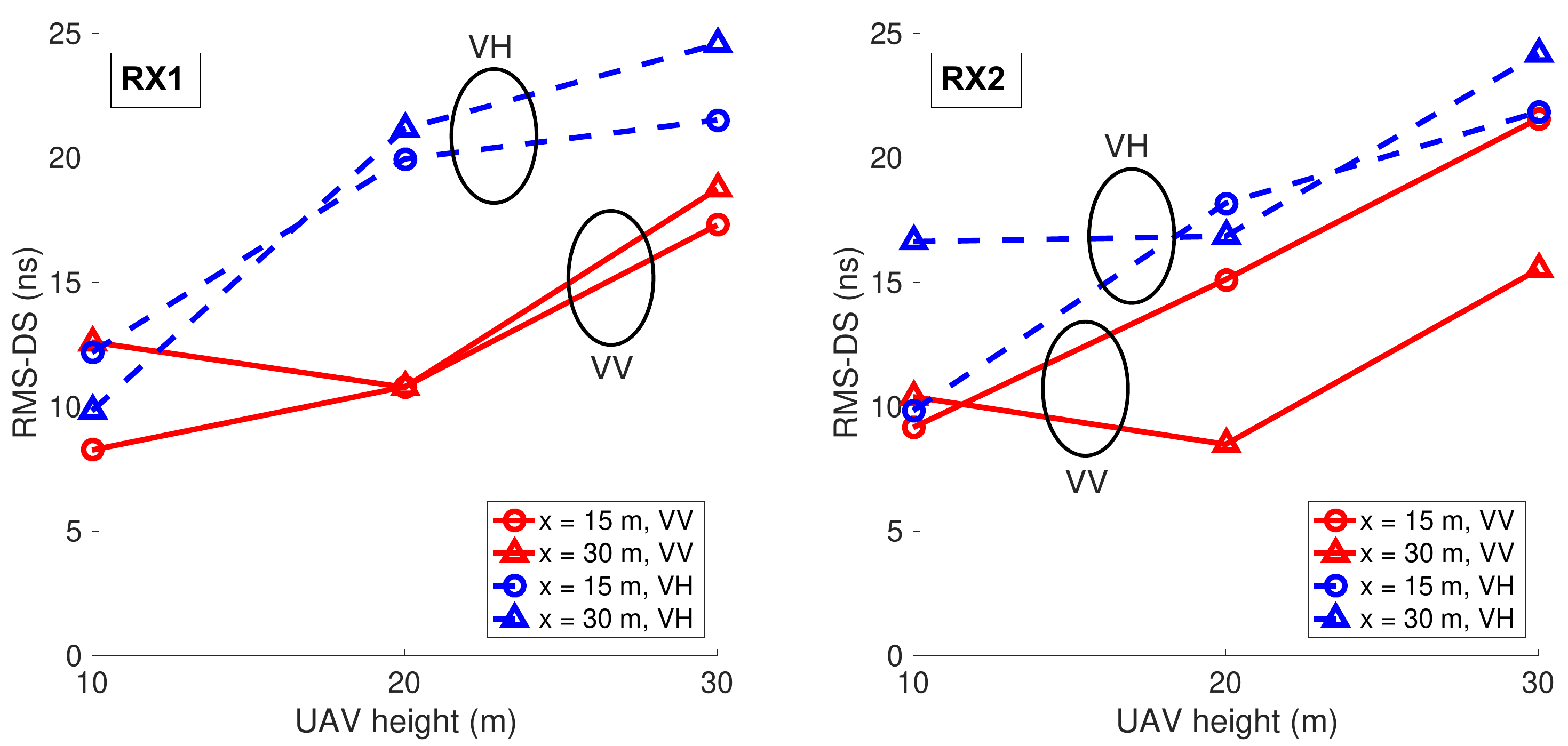}
	 \caption{Foliage obstructed UAV hovering.}
     \end{subfigure}
     	\begin{subfigure}{0.75\textwidth}
	\centering
    \includegraphics[width=\columnwidth]{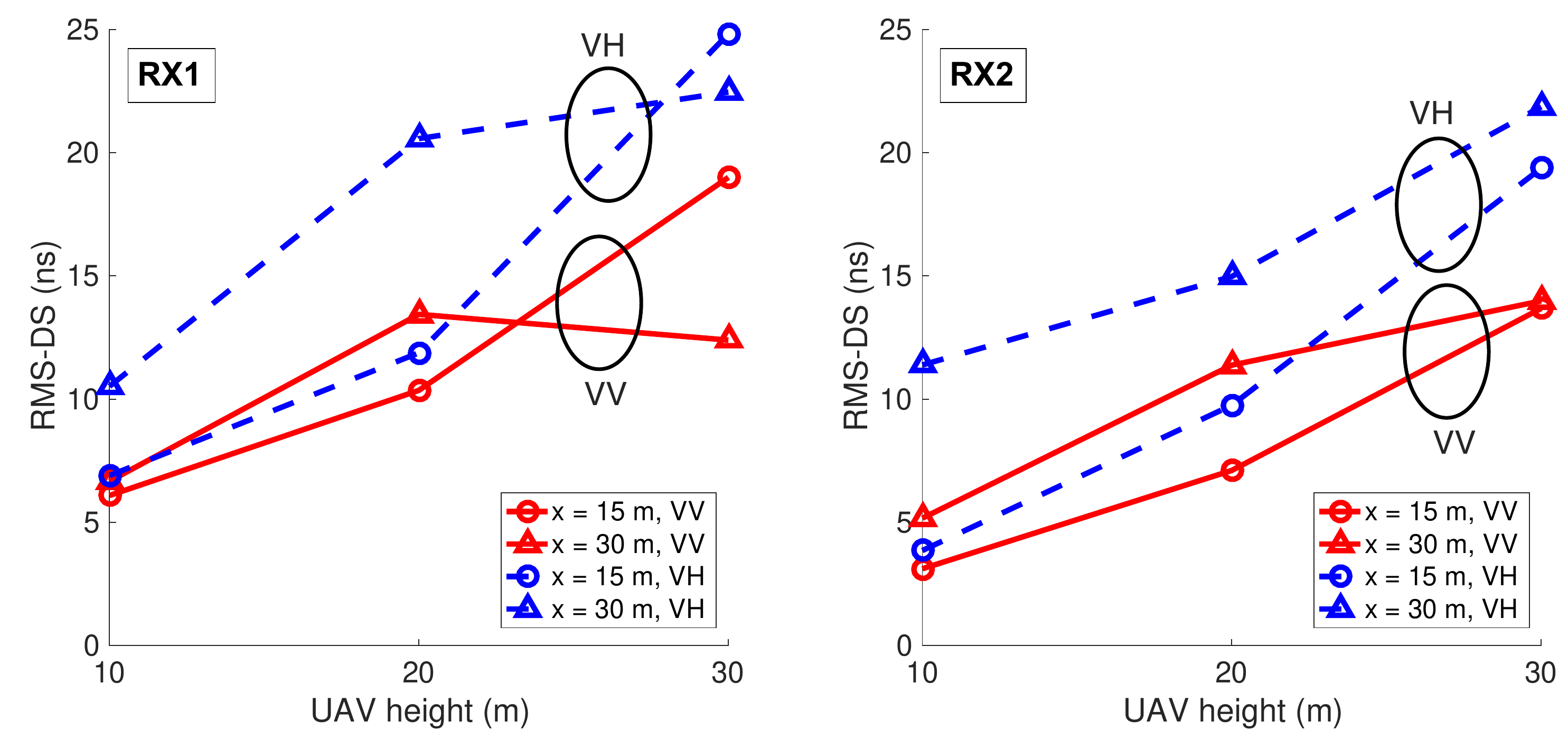}
	 \caption{Unobstructed UAV moving.}
     \end{subfigure}
     \caption{RMS-DS obtained from measurements at RX1 placed on ground~(left) and RX2 at $1.5$~m above the ground~(right) for (a) unobstructed UAV hovering scenario, (b) foliage obstructed UAV hovering scenario, (c) unobstructed UAV moving scenario.}\label{Fig:RMS_DS_mean}
\end{figure}

\subsection{Multipath Channel Analysis: Temporal Dispersion}
Temporal dispersion of the propagation channel is quantified using the RMS-DS represented as $\tau_{\rm rms}$. The RMS-DS is computed using the PDP as follows \cite{wahab_uwb}:
\begin{align}
\tau_{\rm rms} &= \sqrt{\frac{\sum_{\forall n}^{} (n T_{\rm s})^2 \big|H(n)\big|^2}{\sum_{\forall n}^{}\big|H(n)\big|^2}-\Bigg(\frac{\sum_{\forall n}^{} n T_{\rm s} \big|H(n)\big|^2}{\sum_{\forall n}^{}\big|H(n)\big|^2}\Bigg)^2},
\end{align}
where the second term represents the square of the mean delay and $T_{\rm s}$ is the sampling duration~($0.06$~ns). The plot of RMS-DS averaged over the channel scans is shown in Fig.~\ref{Fig:RMS_DS_mean}. A general trend observed is that the RMS-DS increases with the height of the UAV (increases with link distance). Overall, we observe a slightly larger mean RMS-DS at RX1 on the ground than RX2 at 1.5 m height, for all three propagation scenarios for both VV and VH antenna orientations. Also, larger RMS-DS is observed for the VH antenna orientation than the VV antenna orientation at both receiver locations and all three propagation scenarios. As noted, this can be attributed to a weaker dominant component in the VH case as observed in Section~\ref{Section:Significant_MPCs}. 


Overall, for the VV antenna orientation at RX1 and RX2, we observe larger RMS-DS for the unobstructed UAV moving scenario than for the unobstructed UAV hovering scenario shown in Fig.~\ref{Fig:RMS_DS_mean}(c) and (a), respectively. This is likely due to a larger number of MPCs at larger delays arising from different parts of the nearby scatterers during motion. Similarly, for the foliage obstructed scenario shown in Fig.~\ref{Fig:RMS_DS_mean}(b), we observe the highest mean RMS-DS--as noted, likely due to additional MPCs from the foliage. Contrary to the other two scenarios, we observe a larger mean RMS-DS for the foliage obstructed scenario at RX2 than at RX1, likely because a larger number of MPCs are captured from the foliage with the better ground clearance position~(see Section~\ref{Section:Significant_MPCs}). 

A comparison of the VH antenna orientation with VV antenna orientation reveals that we have a smaller effect of the antenna orientation change on the RMS-DS for foliage obstructed and unobstructed UAV moving scenarios~(as observed in Section~\ref{Section:PL} also). Moreover, similar to the VV antenna orientation, we observe a larger difference in RMS-DS between $x=15$~m and $x=30$~m for RX2 than for RX1 for the unobstructed UAV hovering and unobstructed UAV moving scenarios due to the better ground clearance. 
In addition, for the VH antenna orientation, we observe larger mean RMS-DS for the unobstructed UAV hovering scenario than the unobstructed UAV moving scenario at both RX1 and RX2. This can be attributed to spatial averaging of small powered diffuse MPCs during the motion for VH antenna orientation. This averaging may yield a smaller number of MPCs at respective delays compared to the unobstructed UAV hovering scenario for VH antenna orientation.

\section{Modeling Received Power and Propagation Path Loss} \label{Section: Path_Loss_model}
In this section, we model the antenna gain in the elevation plane, analyze the polarization mismatch losses, and provide an analytical model for the received power and path loss for the unobstructed UAV hovering and moving scenarios. Empirical and analytical path loss results are provided. We also describe our ray tracing simulation setup and simulated path loss results for the unobstructed UAV hovering scenario.

\begin{figure*}[t!]
    \centering
    \begin{subfigure}[t]{0.5\textwidth}
        \centering
        \includegraphics[width =\textwidth]{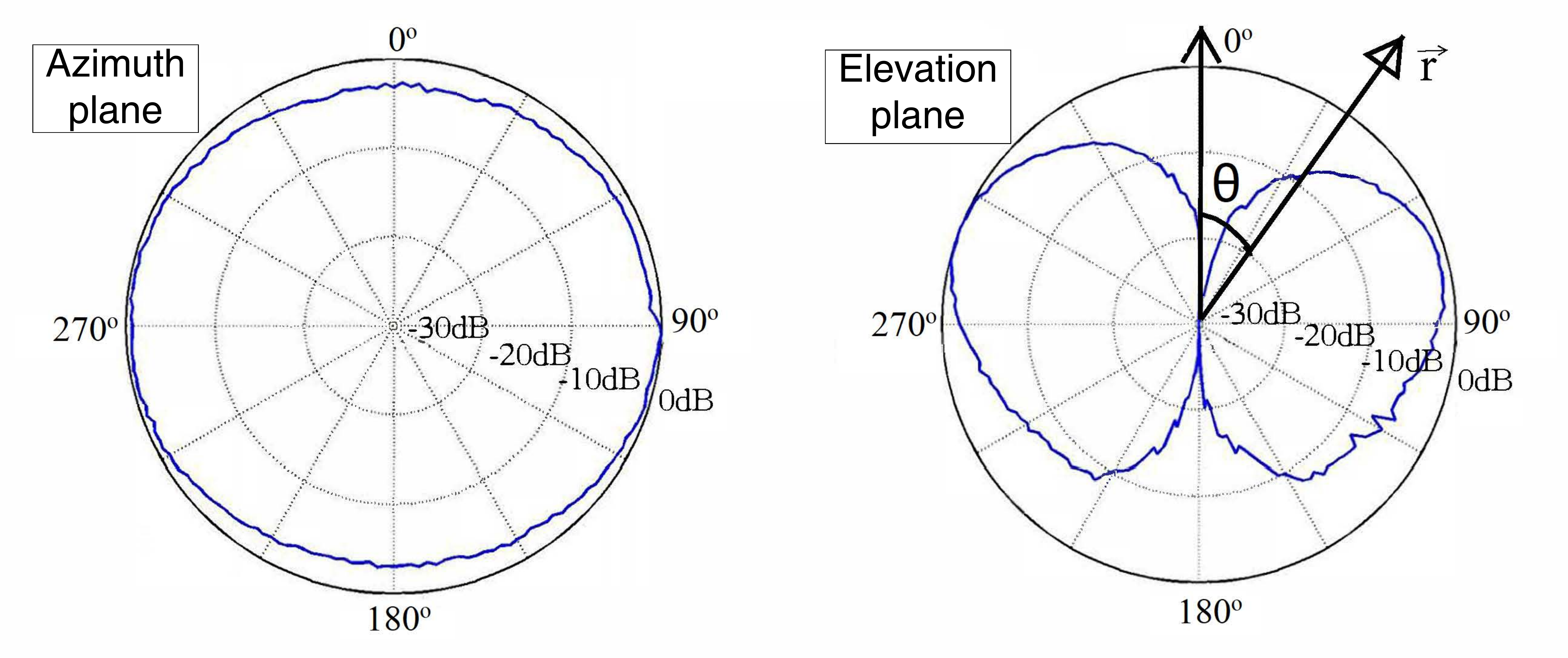}
        \caption{}
    \end{subfigure}%
    ~ 
    \begin{subfigure}[t]{0.5\textwidth}
        \centering
        \includegraphics[width=\textwidth]{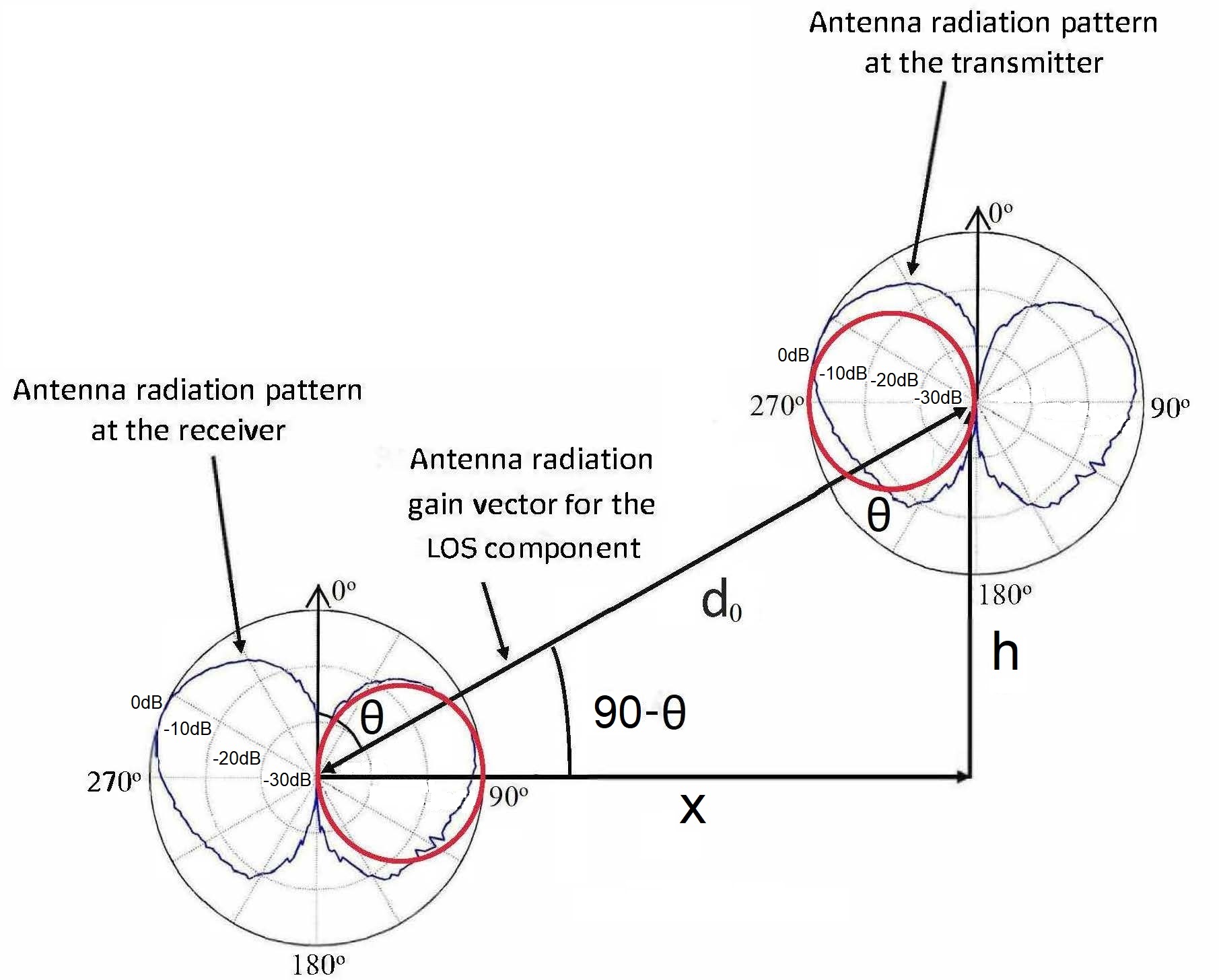}
        \caption{} 
    \end{subfigure}
    \caption{Antenna radiation pattern at $4$~GHz~\cite{Time_Domain}, (a) in the azimuth and elevation planes with directional vector at a given elevation angle, (b) in the elevation plane for the transmitter on the UAV and receiver on the ground station.} \label{Fig:Antenna_pattern}
\end{figure*}

\subsection{Antenna Gain Modeling and Polarization Mismatch Losses} \label{Section:Antenna_model}
BroadSpec UWB antennas from Time Domain Inc.~\cite{Time_Domain} were used in the experiments at both the TX and the RX. These antennas are planar elliptical dipoles with omni-directional pattern in the azimuth plane and a typical doughnut pattern in the vertical plane, shown in Fig.~\ref{Fig:Antenna_pattern}(a). The parameters of the antennas are provided in Table~\ref{Table:Sounding_parameters}. The vector $\vec{r}$ in Fig.~\ref{Fig:Antenna_pattern}(a) represents the direction of the link in the elevation plane at a given elevation angle~$\theta$ given by, $\theta=\tan^{-1}\Big(\frac{x}{h}\Big)$, where $x$ represents the horizontal distance between the RX and the TX, and $h$ represents the height of the UAV. 

\begin{figure}[!t]
	\centering
	\includegraphics[width=0.5\columnwidth]{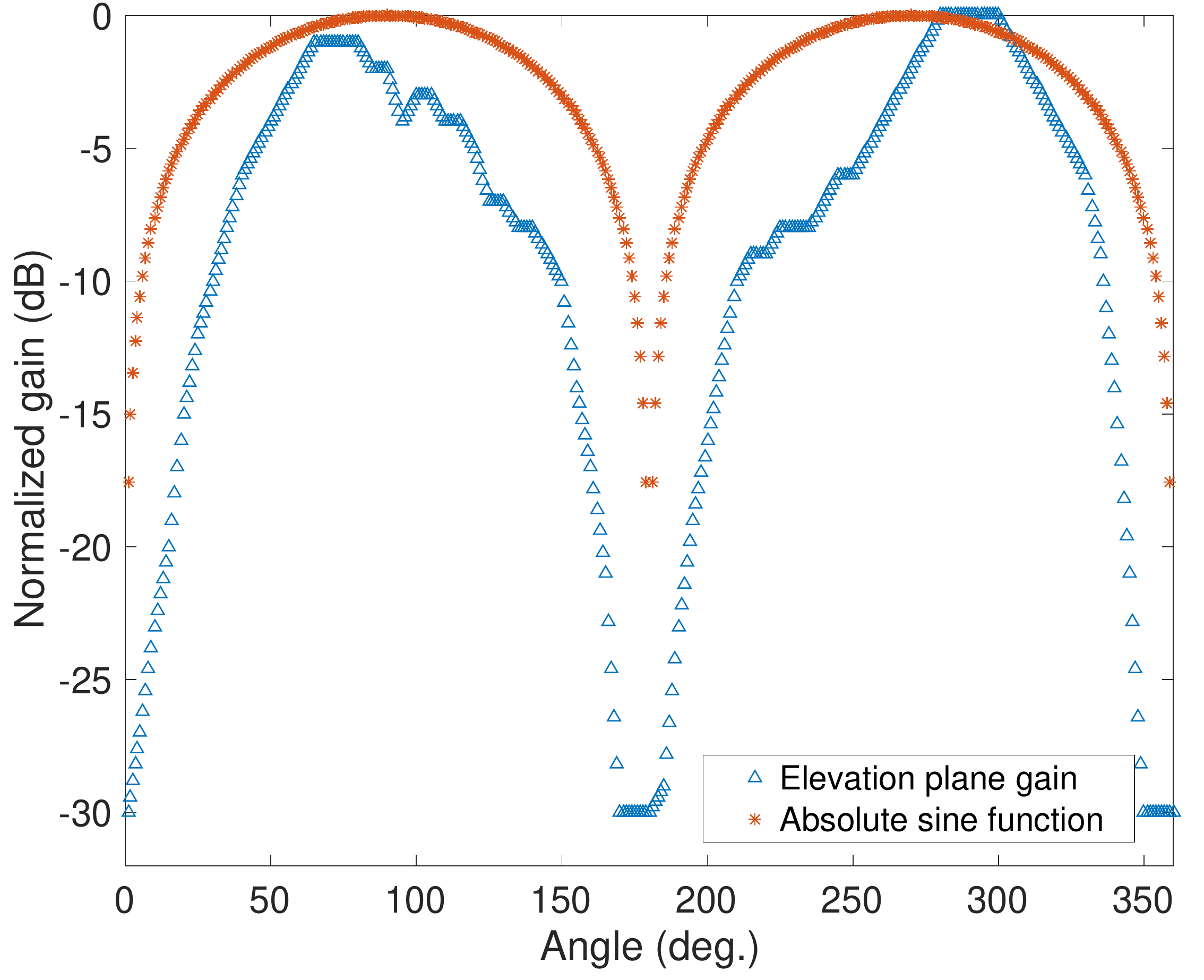}
	\caption{Comparison of normalized antenna gain at $4$~GHz in the elevation plane with the absolute sine function.}\label{Fig:Elevation_gain_sine}
\end{figure}

In AG propagation, it is important to consider the antenna radiation pattern in three dimensions~\cite{wahab_survey}. The antenna radiation pattern in the elevation plane plays a key role in determining the received power, especially at higher elevation angles. The elevation angle is defined between the horizontal and a direct line connecting TX to RX~(along the LOS component, if present). The link in Fig.~\ref{Fig:Antenna_pattern}(b) represents the LOS component from the phase center of the TX antenna to the RX antenna. This LOS component makes an angle $\theta$ with the vertical axis. The antenna radiation pattern is approximately symmetrical around the vertical axis as shown by the red circles in Fig.~\ref{Fig:Antenna_pattern}(b). Therefore, the antenna gain for the LOS component in the elevation plane can be approximated by an absolute sine trigonometric function, i.e., $\sin  \theta$. Comparison of the antenna gain at $4$~GHz in the elevation plane from manufacturer's specification~\cite{Time_Domain} with the absolute sine function $\sin  \theta$ over an angular span of [$0$~$180^\circ$] is provided in Fig.~\ref{Fig:Elevation_gain_sine}. From Fig.~\ref{Fig:Elevation_gain_sine}, we can observe that a sine function with larger exponent value can even provide a better fit. However, as the operating frequency covers a large band, the pattern of the antenna radiation at different frequencies is not the same~(see \cite{jianlin}). Therefore, $\sin \theta$ with larger exponent value does not always provide a good fit. Instead, $\sin \theta$ provides an overall better fit for antenna gain in the elevation plane over the whole frequency range. For the VV antenna orientation, the overall antenna gain for the LOS component can be approximated as $\sin\theta$ over the elevation angle range of $0^{\circ}$ to $180^{\circ}$. Hence, as ${\theta}$ decreases~(an increase of elevation), the LOS component is attenuated accordingly. 

\begin{figure}[!t]
	\centering
	\includegraphics[width=0.5\columnwidth]{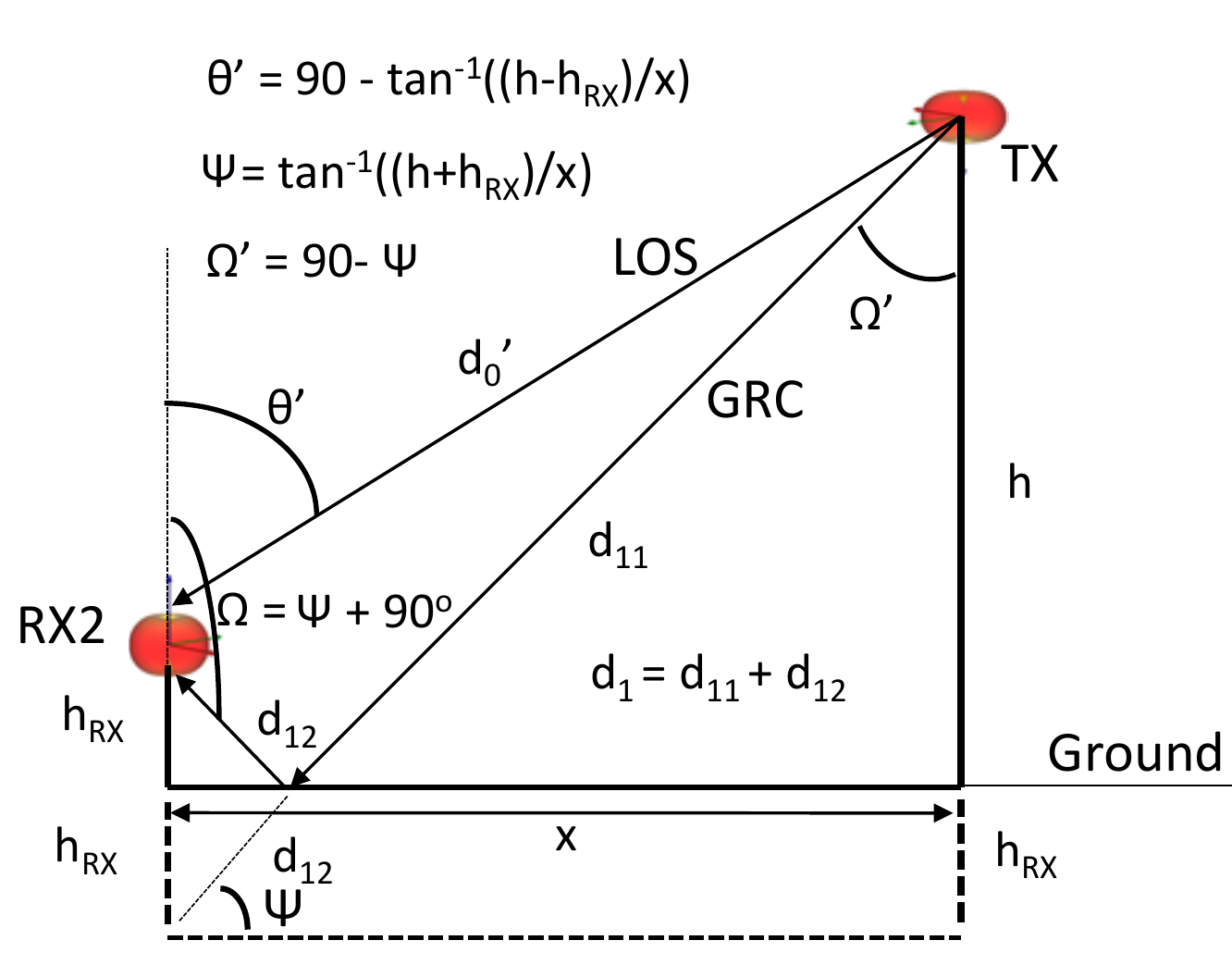}
	\caption{Received signal components of line-of-sight and ground-reflected component at RX2.}\label{Fig:Two_ray_scenario}
\end{figure} 


Similarly, for the ground reflected component~(GRC), the gain of the antenna in the elevation plane is given by $\sin\Omega$ and $\sin\Omega'$ at the RX and TX respectively, from Fig.~\ref{Fig:Two_ray_scenario}. Therefore, the combined TX/RX antenna gain of the GRC  for the VV antenna orientation is given by $\sqrt{\sin\Omega\sin\Omega'}$. On the other hand, for the VH antenna orientation, we have a weak~(or absent) LOS component and most of the propagation is non-line-of-sight~(NLOS),  due to reflections from surrounding objects. 

In order to quantify the polarization mismatch, let $\pmb{\rho}^{\rm (TX)}$ and $\pmb{\rho}^{\rm (RX)}$ represent the unit vectors for the electric fields at the TX and RX antennas, respectively. At the RX antenna, we have $\pmb{\rho}^{\rm (TX)}\cdot\pmb{\rho}^{\rm (RX)}$, where $(\cdot)$ represents the dot product between the two unit vectors. If the direction of the electric field planes~(for incident and receiver) are similar, there is no polarization mismatch and the dot product is $1$. However, for the VH antenna orientation, the incident and received electric field planes will be orthogonal~\cite{antenna_pol,balanis}. In an ideal case, the VH antenna orientation should yield no reception for the linearly polarized antennas. However, due to non-ideal cross-polarization discrimination and due to reflections from scatterers in the environment, cross-polarized components appear. These components enable reception for the VH antenna orientation.

\subsection{Received Power Modeling using LOS and GRC Paths} \label{Section:Rec_pwr}
 If $S(n)$ is the transmitted signal, then the received signal is given by $R(n)=S(n)\circledast H(n)$, where $\circledast$ is the convolution operation and $H(n)$ is as defined in~(\ref{Eq:Eq_CIR2}). If there are $M$ paths in the propagation channel, then the received signal consists of $M$ MPCs, and each MPC can be represented as $R_m(n)$ given by~\cite{two_ray1,two_ray2}:
 \begin{equation}
\begin{split}
        R_m(n)=\bigg[\frac{\lambda \Gamma_m(\phi_m,\theta_m)}{4\pi d_m}\sqrt{G_{\rm T}\bigg(\phi_m^{\rm (TX)},\theta_m^{\rm (TX)}\bigg) G_{\rm R}\bigg(\phi_m^{\rm (RX)},\theta_m^{\rm (RX)}\bigg)}s(n-\tau_m)\exp \bigg(\frac{-j2\pi d_m}{\lambda}\bigg) \Big|\pmb{\rho}^{\rm (TX)}_{m}\cdot\pmb{\rho}^{\rm (RX)}_{m}\Big| \bigg], \label{Eq:Received_comp_overall}
\end{split}
\end{equation}
where $m=0,1,2,\dots,M$ with $m=0$ representing the LOS component, $\lambda$ is the wavelength of the transmitted signal, $\Gamma_m(\phi_m,\theta_m)$ is the reflection coefficient of the $m^{\rm th}$ component with $\phi_m$ and $\theta_m$ being the azimuth and elevation angles of the received components with respect to the scatterers~(considering first order reflections), $G_{\rm T}(\phi_m^{\rm (TX)},\theta_m^{\rm (TX)})$ is the gain of the antenna at the TX at respective azimuth and elevation angles of departures, $G_{\rm R}(\phi_m^{\rm (RX)},\theta_m^{\rm (RX)})$ is the gain of the antenna at the RX at respective azimuth and elevation angles of arrivals, $\Big|\pmb{\rho}^{\rm (TX)}_{m}\cdot\pmb{\rho}^{\rm (RX)}_{m}\Big|$ is the polarization mismatch loss factor for the $m^{\rm th}$ MPC~\cite{balanis}, and finally $\tau_m$ and $d_m$ are the delay and distance of the $m^{\rm th}$ component, respectively. We use the terms reflection and scattering mostly interchangeably here, understanding that these represent distinct physical propagation mechanisms; their aggregate effect is captured by $\Gamma$ in our formulation.

For the LOS component $\Gamma_0(\phi_0,\theta_0) = 1$, and the distance of the path between the TX and RX will be $d_0$, shown in Fig.~\ref{Fig:Antenna_pattern}(b). Similarly, $\tau_0 = 0$ for the LOS component in our case. As noted the gain of the antenna for the LOS component can be approximated by the trigonometric function described in Section~\ref{Section:Antenna_model}. Therefore, the LOS received component can be represented as follows:
\begin{equation}
    \begin{split}
    \begin{aligned}
      R_0(n)&=\bigg[{ \frac{\lambda}{4\pi d_0}\sin\theta s(n)\exp \bigg(\frac{-j2\pi d_0}{\lambda}\bigg)} \Big|\pmb{\rho}^{\rm (TX)}_{0}\cdot\pmb{\rho}^{\rm (RX)}_{0}\Big|\bigg].
       \label{Eq:Received_comp_LOS}
    \end{aligned}
     \end{split}
\end{equation}

Similarly, the GRC can be represented as,
\begin{equation}
    \begin{split}
    \begin{aligned}
R_1(n)&=\bigg[{ \frac{\lambda \Gamma_1(\phi_1,\theta_1)}{4\pi d_1}\sqrt{\sin\Omega\sin\Omega'}s(n-\tau_{1})\exp \bigg(\frac{-j2\pi d_1}{\lambda}\bigg)}\Big|\pmb{\rho}^{\rm (TX)}_{1}\cdot\pmb{\rho}^{\rm (RX)}_{1}\Big| \bigg]. \label{Eq:Received_comp_GRC}
    \end{aligned}
     \end{split}
\end{equation}
If $E$ represents the average over time, the total received power $P_{\rm R}$, as the sum of the powers of all the received MPCs is given as:
  \begin{equation}
 \begin{aligned}
  P_{\rm R}={\rm E}\bigg[\big|R_0(n)\big|^2\bigg]+ {\rm E}\bigg[\sum_{m=1}^{M-1} \big| R_m(n)\big|^2 \bigg], \label{Eq:Total_power} 
 \end{aligned}
 \end{equation}
 where $m=1$ refers to the GRC. 
 
  
 
 \subsection{Path Loss Modeling for Unobstructed UAV Hovering Scenario, VV} \label{Section:PL_modeling_hovering}
 The path loss is obtained by comparing the received power at a given distance with the received power at the reference distance. First, the received power is obtained at $1$~m distance when the antennas are aligned boresight to each other and at the same height of $1$~m above the ground. Then, the received power at different distances is compared with that at the reference distance. Let us first consider the unobstructed UAV hovering scenario with VV antenna orientation at RX1 and RX2 shown in Fig.~\ref{Fig:Scenario_AG}(a). For RX1, we consider that the received power is mainly from the dominant LOS component for different UAV distances as discussed in Section~\ref{Section:Rec_pwr}. Moreover, for simplicity, we consider the polarization mismatch loss factor from (\ref{Eq:Received_comp_LOS}) equal to $1$ for the LOS component. Therefore, from (\ref{Eq:Received_comp_LOS}) and (\ref{Eq:Total_power}), the received power for the LOS components $P_{{\rm R},d_0}^{\rm (LOS)}$ at different UAV distances can be written as, 
 \begin{align}
     P_{{\rm R},d_0}^{\rm (LOS)} = \frac{P_{\rm T}\sin^2\theta\lambda^2}{\big(4\pi d_{0}\big)^2}, \label{Eq:Rec_power_LOS}
    \end{align}
whereas the received power $P_{\rm R, \textit{d}_{\rm ref}}^{\rm (LOS)}$ at a reference distance $d_{\rm ref}$ = $1$~m, can be written as:
\begin{align}
     P_{\rm R, \textit{d}_{\rm ref}}^{\rm (LOS)} = \frac{P_{\rm T}\lambda^2}{\big(4\pi \big)^2},
 \end{align}
 where $P_{\rm T} = {\rm E}\big[|s(n)|^2\big]$ is the transmitted power, $\theta = \tan^{-1}\big(\frac{x}{h}\big)$, and distance $d_{0}= \sqrt{x^2 + h^2}$ is from Fig.~\ref{Fig:Antenna_pattern}(b). Additionally, the antenna gains at the TX and RX sides at a reference distance of $1$~m are approximated as $1$ ($0$~dB) which is the maximum antenna gain in the far field at the boresight alignment of the antennas. Using the close-in free space path loss model~\cite{rappaport}, the path loss~(in dB) can be represented as: 
\begin{equation}
L(d)~[\rm{dB}]=10\log_{10}{L(\textit{d}_{\rm ref})} + 10\log_{10} \frac{P_{\rm R, \textit{d}_{\rm ref}}}{P_{{\rm R},\textit{d}}}, \label{Eq:Path_loss_main}
\end{equation}
where $10\log_{10}L(d_{\rm ref})$ is the free space path loss~(in dB) at the reference distance, $L(\textit{d}_{\rm ref}) = (\frac{4\pi \textit{d}_{\rm ref}}{\lambda})^2$~\cite{rappaport}, and $\lambda$ corresponds to the wavelength at the center of the UWB signal spectrum. Therefore, at RX1, the path loss for VV antenna orientation can be approximated based on the LOS component only (\ref{Eq:Rec_power_LOS}) as follows: 
\begin{equation}
\begin{split}
    L^{(\rm VV)} (d_0)~[\rm{dB}]&=10\log_{10}L(d_{\rm ref}) + 10\log_{10}\frac{d_{0}^2}{\sin^2\theta}\\ &=10\log_{10}L(d_{\rm ref}) + 10\log_{10}\frac{x^2 + h^2}{\Bigg[\sin\bigg(\tan^{-1}\big(\frac{x}{h}\big)\bigg)\Bigg]^2}. \label{Eq:Path_loss_RX1}
\end{split}
\end{equation}

The path loss at RX2 is calculated in a similar way. However, at RX2, in addition to the LOS component, we have a dominant GRC due to the height of the RX above the ground as shown in Fig.~\ref{Fig:Two_ray_scenario}. Due to the large temporal resolution of the transmitted signal, the GRC and the LOS component can be resolved. Therefore, from (\ref{Eq:Received_comp_LOS}) and (\ref{Eq:Received_comp_GRC}), the overall received power can be approximated as: \begin{equation}
\begin{split}
   P_{\rm R} =& P_{\rm R}^{\rm (LOS)} + P_{\rm R}^{\rm (GRC)} \\
    P_{\rm R} =& \frac{P_{\rm T} G_{\rm T}^{\rm (LOS)}(\phi_0^{\rm (TX)},\theta_0^{\rm (TX)}) G_{\rm R}^{\rm (LOS)}(\phi_0^{\rm (RX)},\theta_0^{\rm (RX)})\lambda^2}{(4\pi d_{0}')^2}~+\\
    &\frac{P_{\rm T} G_{\rm T}^{\rm (GRC)}(\phi_1^{\rm (TX)},\theta_1^{\rm (TX)}) G_{\rm R}^{\rm (GRC)}(\phi_1^{\rm (RX)},\theta_1^{\rm (RX)})\lambda^2|\Gamma_1(\phi_1,\theta_1)|^2}{(4\pi d_{1})^2}\Big|\pmb{\rho}^{(\rm{TX})}_{1}\cdot\pmb{\rho}^{(\rm{RX})}_{1}\Big|^2,
\end{split}
\end{equation}
where from Fig.~\ref{Fig:Two_ray_scenario}, we have $d_{0}' = \sqrt{(h-h_{\rm RX})^2 + x^2}$, $d_{1} = \sqrt{(h + h_{\rm RX})^2 + x^2}$, and $\Gamma_1(\phi_1,\theta_1) = \Gamma_1(\Psi)$. The elevation angle of the GRC at the TX and RX is represented as $\Omega'$ and $\Omega$, respectively. The angle of the LOS component is also modified as $\theta'$ due to the height of the receiver above the ground. Moreover, the value of 
$\Big|\pmb{\rho}^{(\rm{TX})}_{1}\cdot\pmb{\rho}^{(\rm{RX})}_{1}\Big|^2$ is approximated as 1, similar to the LOS component. Therefore, considering only the LOS and GRC paths, the path loss for RX2 from (\ref{Eq:Path_loss_main}) is given as follows:
\begin{equation}
\begin{split}
    L^{(\rm VV)} (d_0',d_1)~[\rm{dB}] = &10\log_{10}L(d_{\rm ref}) +~ 10\log_{10}\frac{(d_{0}'d_{1})^2}{(d_{1}\sin\theta' )^2 + (d_{0}'^2\sin\Omega\sin\Omega')|\Gamma_1(\Psi)|^2}. \label{Eq:Hovering_RX2}
\end{split}
\end{equation}
These analytical path loss results at RX1 and RX2 will be compared with measurements in Section~\ref{Section:PL}. 


\subsection{Path Loss Modeling for Unobstructed UAV Moving Scenario, VV} \label{Section:PL_modeling_moving}
The path loss for the unobstructed UAV moving scenario with VV antenna orientation can be modeled in a similar way as for the unobstructed UAV hovering scenario. However, the effect of the antenna gain is different at the TX and the RX. At the TX, mounted on the UAV, the antenna's boresight during motion is fixed with respect to the RXs on the ground. Therefore, the TX antenna gain can be modeled with $\sin\theta$, same as for the UAV hovering scenario. On the other hand, the gain of the RX antennas on the ground is not fixed during the UAV motion. This is because the gain of the antenna is not the same at a given elevation point in space around the antenna. However, due to circular motion around the RX, the overall RX antenna gain can be approximated with a mean value represented as, $G_{\rm R}^{(\rm {c})}$. Therefore, the received power for the LOS component at RX1 for UAV moving scenario is given as:
\begin{align}
     P_{{\rm R},d_0}^{\rm (LOS)} = \frac{P_{\rm T}\sin\theta G_{\rm R}^{(\rm {c})}\lambda^2}{\big(4\pi d_{0}\big)^2}.~
     \label{Eq:Moving1}
 \end{align}
 The path loss at RX1 when the UAV is moving can therefore be represented as:
 \begin{equation}
\begin{split}
    L^{(\rm VV)} (d_0)~[\rm{dB}] = 10\log_{10}L(d_{\rm ref}) + 10\log_{10}\frac{d_0^2}{\sin\bigg(\tan^{-1}\big(\frac{x}{h}\big)\bigg)G_{\rm R}^{(\rm {c})}}. \label{Eq:Pathloss_moving1}
\end{split}
\end{equation}

Similar to RX1, the path loss at RX2 when the UAV is moving can be represented as
\begin{equation} 
\begin{split}
    &L^{(\rm VV)} (d_0',d_1)~[\rm{dB}] = \\ &10\log_{10}L(d_{\rm ref}) +~ 10\log_{10}\frac{(d_{0}'d_{1})^2}{d_{1}^2\sin\theta'G_{\rm R}^{(\rm {c})} + d_{0}'^2\sin\Omega'G_{\rm R}^{(\rm {c})}|\Gamma_1(\Psi)|^2}. \label{Eq:Pathloss_moving2}
\end{split}
\end{equation}
These analytical results will be compared with measurements in Section~\ref{Section:PL}.

 \begin{figure}[!t]
    \centering
	\begin{subfigure}{0.7\textwidth} 
	\centering
	\includegraphics[width=\columnwidth]{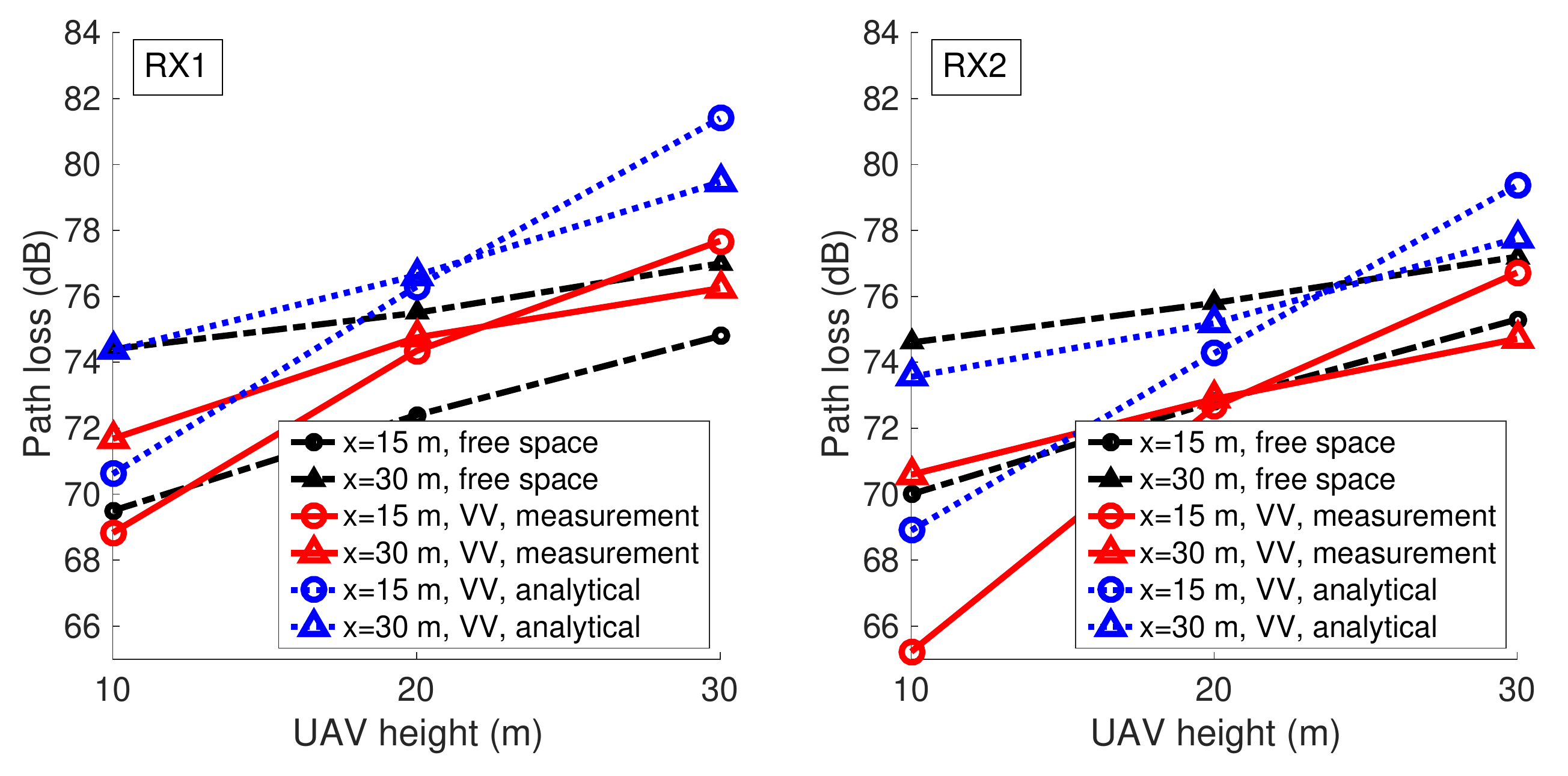} 
	\centering
	\caption{Unobstructed UAV hovering.}
	 \end{subfigure}
	\begin{subfigure}{0.70\textwidth}
	\centering
    \includegraphics[width=\columnwidth]{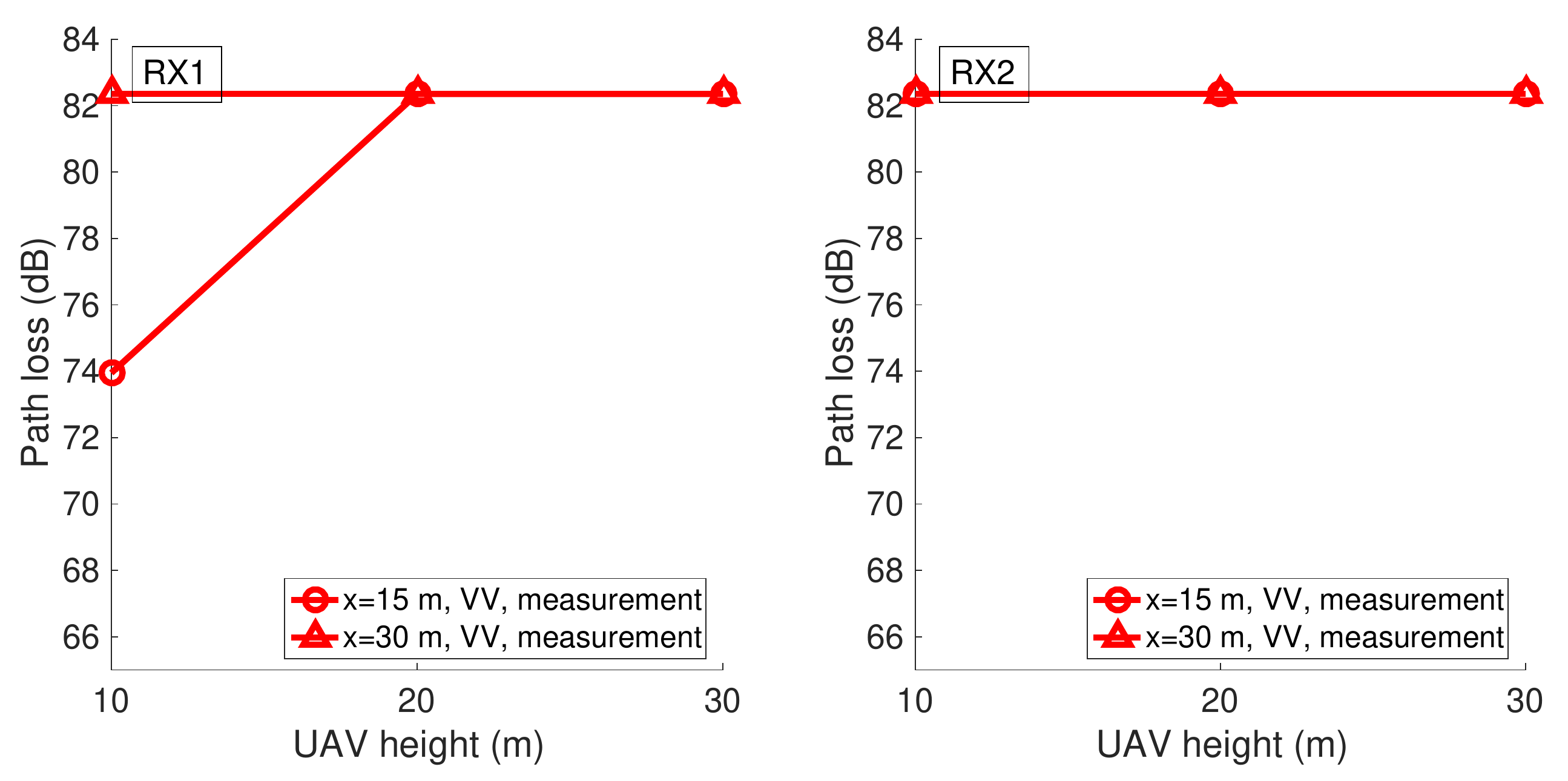}
	 \caption{Foliage obstructed UAV hovering.}
     \end{subfigure}
     	\begin{subfigure}{0.70\textwidth}
	\centering
	\vspace{-1mm}
    \includegraphics[width=\columnwidth]{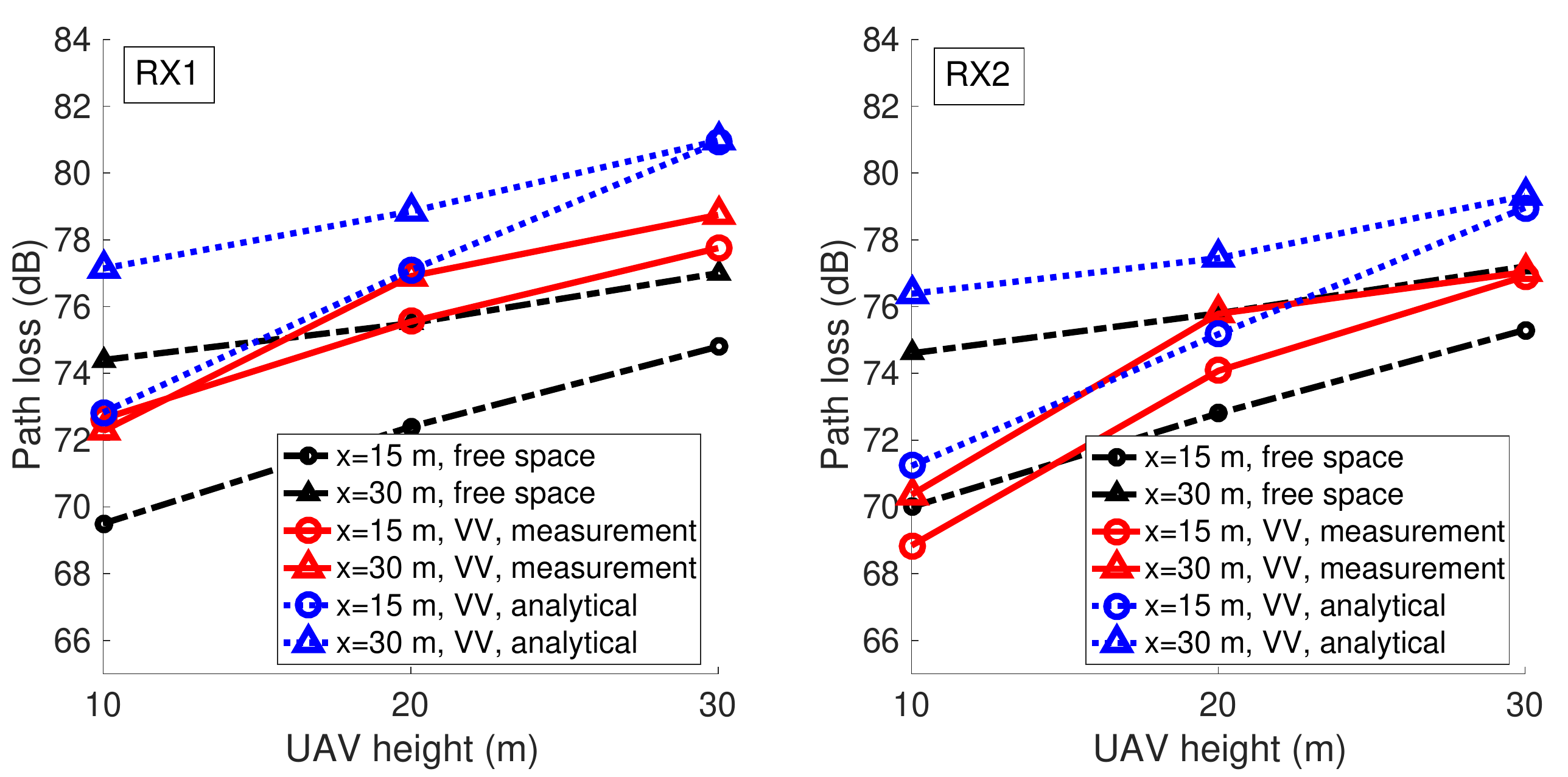}
	 \vspace{-4mm}
	 \caption{Unobstructed UAV moving.}
     \vspace{-1mm}
     \end{subfigure}
     \vspace{-2mm}
     \caption{Path loss obtained from free space, measurements and analytical modeling for VV antenna orientation at RX1 placed on ground~(left) and RX2 at $1.5$~m above the ground~(right) for (a) unobstructed UAV hovering, (c) unobstructed UAV moving. The measurement results for UAV hovering with foliage obstruction are shown in (b).}\label{Fig:PL_mean_VV}
     \vspace{-1mm}
\end{figure}

The path loss for the foliage scenario remains essentially constant for different UAV heights and horizontal distances as the LOS path is blocked by the foliage. There are small variations of the received power~(due to weak MPCs) versus link distances, however, the logarithm smooths these variations resulting in an approximately constant path loss. The reason for this constant value is due to the limitation of the equipment to measure the minimum received power during obstruction, i.e., a limited dynamic range.

\begin{figure}[!t]
    \centering
	\begin{subfigure}{0.7\textwidth} 
	\centering
	\includegraphics[width=\columnwidth]{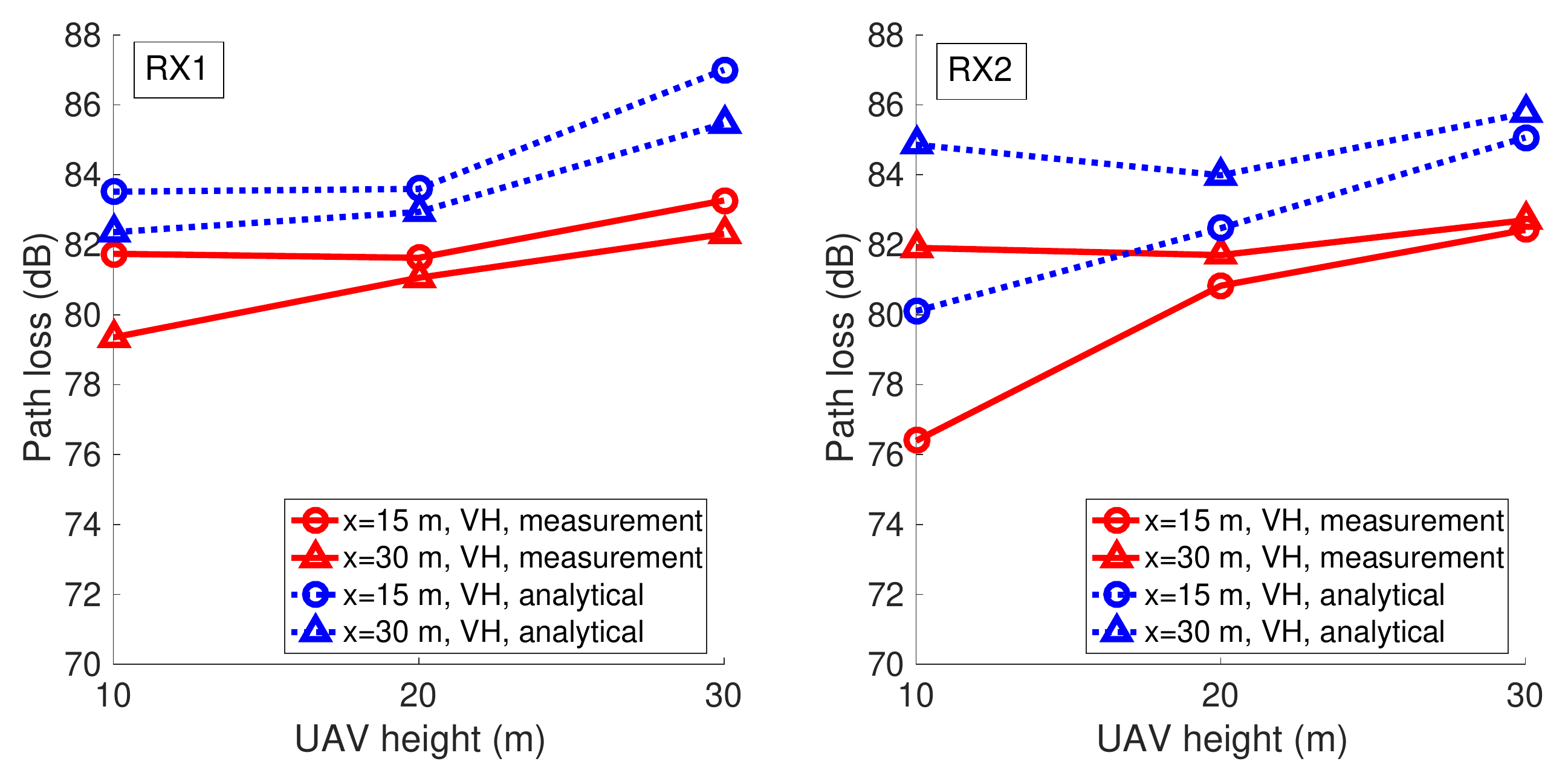} 
	\centering
	\caption{Unobstructed UAV hovering.}
    \end{subfigure}			
	\begin{subfigure}{0.7\textwidth}
	\centering
    \includegraphics[width=\columnwidth]{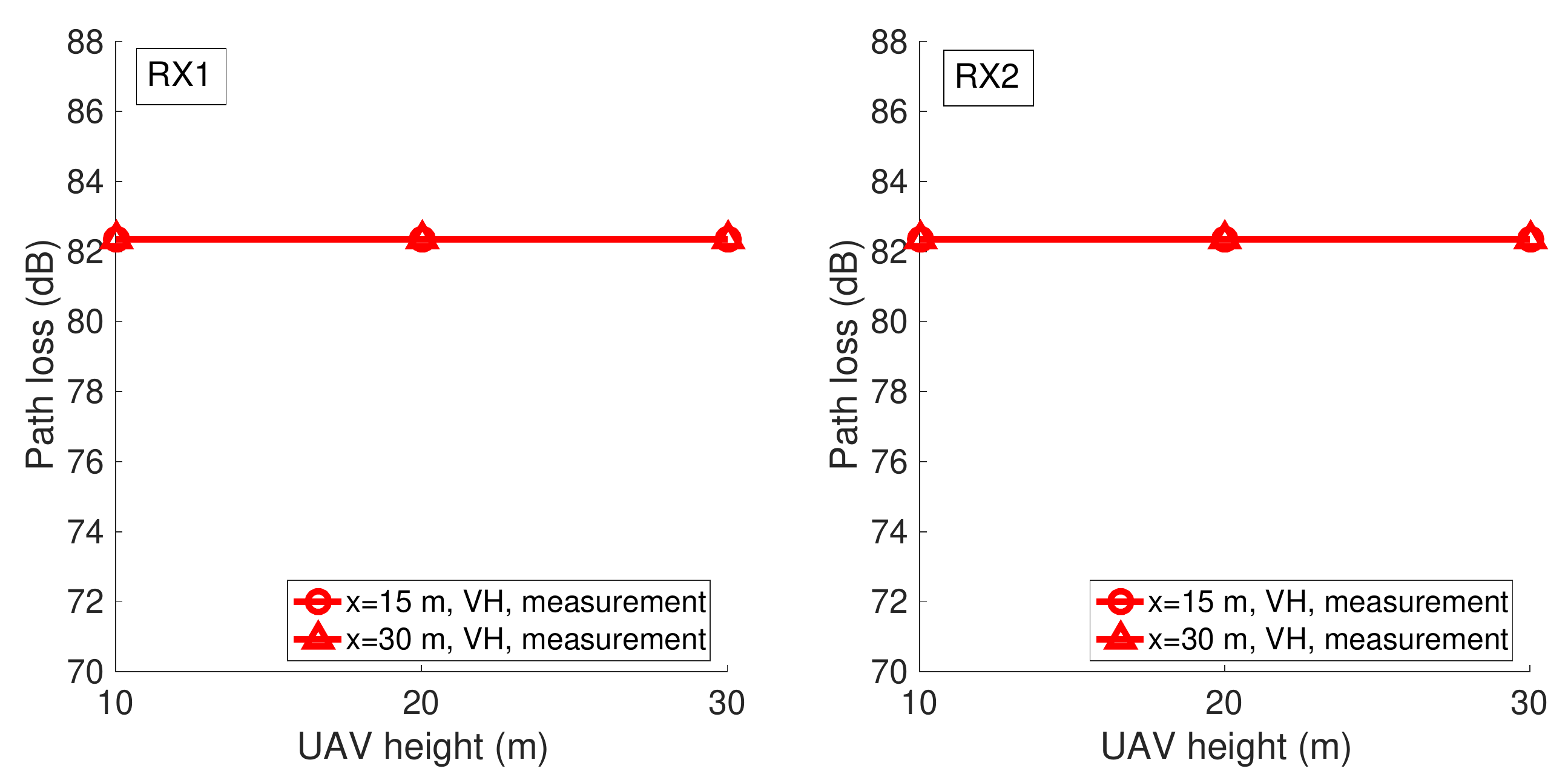}
	 \caption{Foliage obstructed UAV hovering.}
     \end{subfigure}
     	\begin{subfigure}{0.7\textwidth}
	\centering
    \includegraphics[width=\columnwidth]{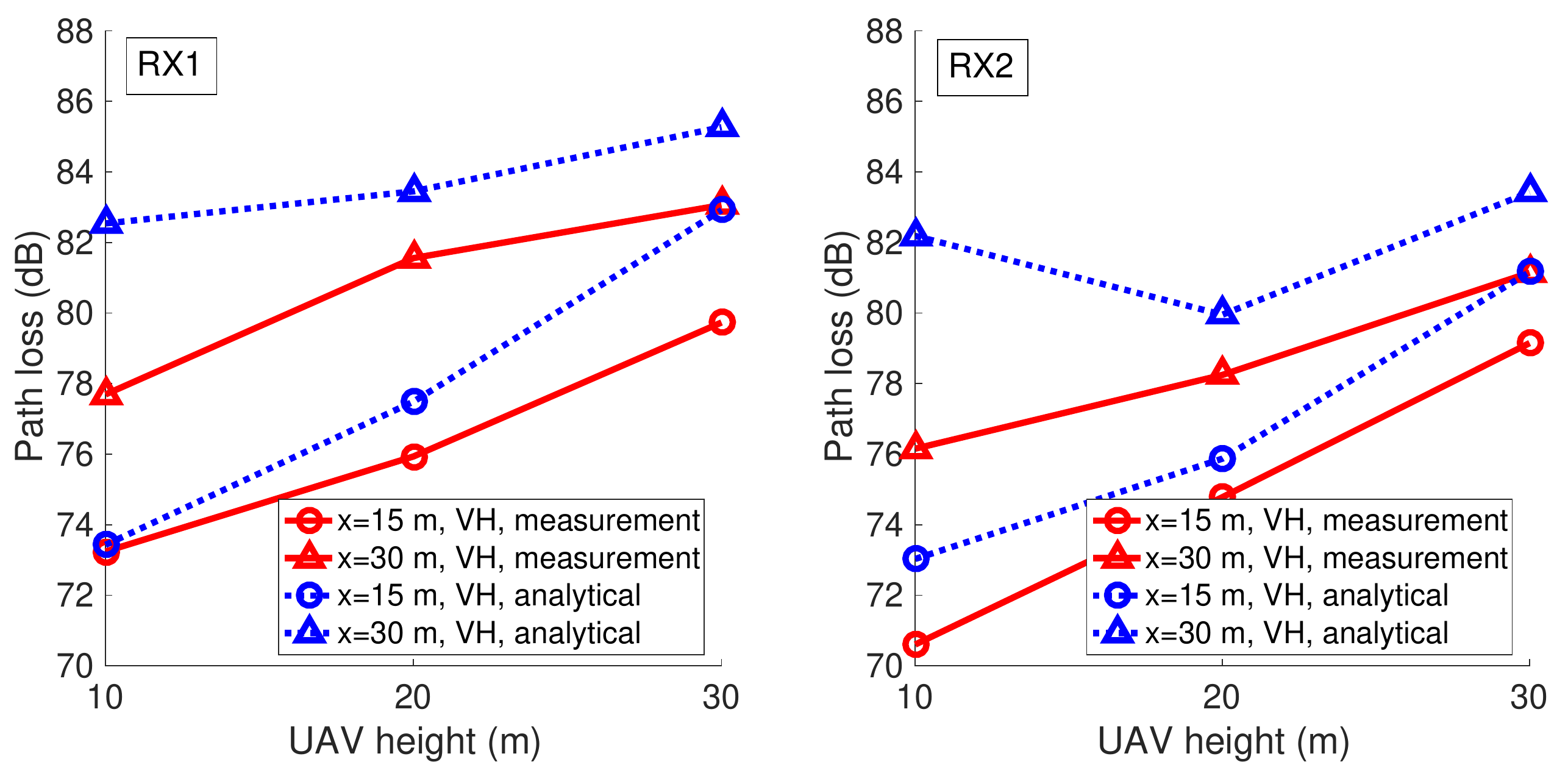}
	 \caption{Unobstructed UAV moving.}
     \end{subfigure}
     \caption{Path loss obtained from free space, measurements and analytical modeling for VH antenna orientation at RX1 placed on ground~(left) and RX2 at $1.5$~m above the ground~(right) for (a) unobstructed UAV hovering, (c) unobstructed UAV moving. The measurement results for UAV hovering with foliage obstruction are shown in (b). } \label{Fig:PL_mean_VH}
\end{figure}

\subsection{Path Loss Modeling for Unobstructed UAV Hovering and Moving Scenarios, VH} \label{Section:PL_modeling_VH}
The polarization mismatch loss for the VH antenna orientation can be estimated empirically. In order to obtain this estimate, the received power measurements with VH antenna orientation are compared with the respective measurements with VV antenna orientation. Since the channel is assumed to be stationary and the received power is mainly dependent on the dominant LOS component, from (\ref{Eq:Received_comp_LOS}) and (\ref{Eq:Total_power}), the only variable quantity for VH antenna orientation compared to VV antenna orientation is ($\pmb{\rho}^{\rm (TX)}_{0}\cdot\pmb{\rho}^{\rm (RX)}_{0}$). The polarization mismatch loss factor ratio $c_{\rm pol}$ between the VH and VV antenna orientation scenarios can be represented as:
\begin{equation}
    c^{}_{\rm pol} = \frac{|\pmb{\rho}^{\rm (TX,VH)}_{0}\cdot\pmb{\rho}^{\rm (RX,VH)}_{0}|^2}{|\pmb{\rho}^{\rm (TX,VV)}_{0}\cdot\pmb{\rho}^{\rm (RX,VV)}_{0}|^2}, \label{Eq:Polar_loss_factor}
\end{equation}
where $\pmb{\rho}^{\rm (TX,VH)}_{0},~\pmb{\rho}^{\rm (TX,VV)}_{0},~\pmb{\rho}^{\rm (RX,VH)}_{0},~\pmb{\rho}^{\rm (RX,VV)}_{0}$ are the polarization unit vectors for VH and VV antenna orientations at the TX and the RX, respectively. Note that we have already approximated $|{\pmb{\rho}^{\rm (TX,VV)}_{0}\cdot\pmb{\rho}^{\rm (RX,VV)}_{0}}|^2 = 1$, meaning that there is no polarization mismatch for the LOS component. Therefore, we can write~(\ref{Eq:Polar_loss_factor}) as, $c_{\rm pol} = |\pmb{\rho}^{\rm (TX,VH)}_{0}\cdot\pmb{\rho}^{\rm (RX,VH)}_{0}|^2$ and this provides an estimate of the overall reduction in the received power~(or increase in the path loss) when the orientation of the antennas is changed from VV to VH.

Using (\ref{Eq:Polar_loss_factor}), we can find the analytical path loss for the UAV hovering and moving scenarios with VH antenna orientation. If $c_{\rm pol}$ represents the polarization mismatch loss for a given VH antenna orientation scenario, then the corresponding path loss is given as
\begin{align}
    L^{(\rm {VH})}(d_0)~[\rm {dB}] = L^{(\rm {VV})}(d_0)~[\rm {dB}] + 10\log_{10}(c_{\rm pol}).
\end{align}
The path loss at RX2 is obtained in a similar way. These results will be compared with measurements in Section~\ref{Section:PL}.

\subsection{Empirical and Analytical Path Loss Results} \label{Section:PL}
In this subsection, we first provide empirical path loss results, shown in Fig.~\ref{Fig:PL_mean_VV} and Fig.~\ref{Fig:PL_mean_VH} for VV and VH antenna orientations, respectively. Subsequently we compare the analytical results from Section~\ref{Section:PL_modeling_hovering}, \ref{Section:PL_modeling_moving} and \ref{Section:PL_modeling_VH} with measurement results. Free space path loss calculated at the center frequency of $3.95$~GHz for co-polarization is also provided for comparison.

Fig.~\ref{Fig:PL_mean_VV} shows the path loss for the three propagation scenarios at RX1 and RX2 for VV antenna orientation. At RX1, for the unobstructed UAV hovering scenario shown in Fig.~\ref{Fig:PL_mean_VV}(a), we observe higher path loss at UAV height of $10$~m for $x=30$~m than for $x=15$~m due to larger link distance. However, as the UAV height increases, the path loss at $x=15$~m increases faster than that at $x=30$~m. As a result, the path loss at UAV height of $30$~m becomes larger for the UAV at $x=15$~m than at $x=30$~m. This increase in the path loss is due to the smaller antenna gain at higher elevation angles as discussed in Section~\ref{Section:Antenna_model}, and results in the two curves for VV crossing each other as the UAV height increases. Similar observations were made at RX2 for this scenario. This is a critical observation, since larger path loss can be observed at shorter distance due to antenna effects and TX/RX geometry.

The path loss for unobstructed UAV moving scenario and VV antenna orientation is shown Fig.~\ref{Fig:PL_mean_VV}(c). Here we observe closely spaced curves. This can be explained by considering the antenna gain in the azimuth plane, shown in Fig.~\ref{Fig:Antenna_pattern}(a). Due to the motion of the UAV in a circular path, the RX antenna gain will change continuously in the azimuth and elevation planes. In the azimuth plane, the gain of the antenna is smaller at $0^\circ$ and $180^\circ$. Similarly, the antenna gain in the elevation plane has slightly larger values at certain elevation angles, as can be observed in Fig.~\ref{Fig:Elevation_gain_sine}. However, considering a large number of samples of the antenna gain pattern in the elevation plane forming the overall three dimensional pattern, the antenna gain variations in the elevation plane are somewhat averaged out by the UAV motion. This phenomenon leads to closely spaced path loss curves for unobstructed UAV moving scenario at $x=15$~m and $x=30$~m for three UAV heights. Similar observations can be made at RX2.

The path loss results for VH antenna orientation are shown in Fig.~\ref{Fig:PL_mean_VH} for RX1 and RX2. The path loss here is larger than the VV antenna orientation at RX1 and RX2 due to polarization mismatch as discussed in Section~\ref{Section:Antenna_model}. Hence the effect of the antenna gain in the elevation plane is negligible compared to that in the VV antenna orientation. An interesting observation is that the polarization mismatch has a smaller effect for the unobstructed UAV moving scenario than for the unobstructed UAV hovering scenario. This is due to a larger number of cross-polarized components arising during the UAV motion than when the UAV is hovering, and also suppression of the dominant LOS component. For VH antenna orientation, the boresight of the antenna is facing the ground~(direction of ``azimuth" emission) shown in Fig.~\ref{Fig:Scenario_AG}. Thus the ground and any ground-based scattering objects are illuminated and additional components reach the RXs. The motion of the UAV results in more cross-polarized components~(assuming the ground surface is not uniform) than when the UAV is hovering. This results in higher received power. This phenomenon is more evident at $x=15$~m than at $x=30$~m. The weak cross-polarized components generated at $x=15$~m are stronger than at $x=30$~m.

The path loss for unobstructed UAV hovering scenario shows the largest change experienced due to antenna orientation misalignment~VH compared to aligned~VV shown in Fig.~\ref{Fig:PL_mean_VH}(a). In contrast, for the foliage obstructed UAV hovering scenario shown in Fig.~\ref{Fig:PL_mean_VH}(b), the path loss approximately remains constant. This is because of the obstruction of the dominant LOS path which carries the largest power~--~significantly larger~(approximately $20$~dB) than the other MPCs. The path loss for the foliage obstructed scenario does not show any significant effect of antenna orientation or link distance for either RX1 or RX2. The path loss only shows a change at RX1 for UAV height of $10$~m at $x=15$~m. This is likely due to the smaller obstruction~(thin tree trunk) experienced by the RX placed on the ground from the low altitude UAV. However, for RX2, and larger UAV heights, the tree crown obstructs the path. Moreover, at a horizontal distance of $x=30$~m, the visibility of the RXs from the UAV become better, yet, the path loss is larger because of larger link distance. Therefore, path loss remains approximately constant.     

In addition to link distance $d$ between the GS and the UAV, the path loss is dependent on the elevation angle between the TX and the RX~(\ref{Eq:Received_comp_LOS}), and antenna orientation--see~Section~\ref{Section:Antenna_model}. The larger the elevation angle and orientation mismatch, the greater the path loss. This effect is more prominent when the antennas have the same orientation. Moreover, the foliage introduces additional path loss due to partial obstruction of the LOS component. We also note that the path loss at RX1 is larger than at RX2 for the unobstructed UAV hovering and unobstructed UAV moving scenarios for both VV and VH antenna orientations. This is attributable to antenna deformation at RX1. RX2 has better ground clearance than RX1. 


The path loss results obtained using the analytical modeling for unobstructed UAV hovering and moving scenarios~(discussed in Section~\ref{Section:PL_modeling_hovering}, \ref{Section:PL_modeling_moving} and \ref{Section:PL_modeling_VH}) are shown in Fig.~\ref{Fig:PL_mean_VV} and Fig.~\ref{Fig:PL_mean_VH} alongside the measurement results. The value of $|\Gamma_1(\Psi)|$ is obtained from \cite{GRC_calc}, for vertical polarization and relative permittivity of $35$ for the grassy ground surface at respective grazing angles. The value of $|\Gamma_1(\Psi)|$ is provided in Table~\ref{Table:Polarization_loss} and is same for both UAV hovering and moving scenarios. The value of $|c_{\rm pol}|$ from Section~\ref{Section:PL_modeling_VH}, for different propagation scenario settings is also shown in Table~\ref{Table:Polarization_loss}. The value of the receiver antenna gain $G_{\rm R}^{(\rm {c})}$, for unobstructed UAV moving scenario~(discussed in Section~\ref{Section:PL_modeling_moving}) is taken as $0.5$. Comparing the analytical results with the empirical results we observe slightly larger path loss for the analytical results. This is mainly due to considering only the LOS component for RX1, and LOS and GRC for RX2 and ignoring the other MPCs in the analytical modeling. Overall, we observe a reasonably close match between the empirical results and our analytical results that incorporate $3$D antenna radiation effects. 

		
        
        
		

\begin{table*}[!h]
\centering
\begin{tabular}{|p{.07cm}|p{0.07cm}|p{0.07cm}|p{0.07cm}|p{0.07cm}|p{0.07cm}|p{0.07cm}|p{0.07cm}|p{0.07cm}|p{0.07cm}|p{0.07cm}|p{0.07cm}|p{0.07cm}|}
\hline
		\multicolumn{1}{|p{.3cm}|}{}&\multicolumn{3}{|p{2.5cm}|}{\textbf{~~RX1~(VV), x = 15~m}}&\multicolumn{3}{|p{2.5cm}|}{\textbf{~~RX1~(VV), x = 30~m}}&\multicolumn{3}{|p{2.5cm}|}{\textbf{~~RX2~(VV), x = 15~m}}&\multicolumn{3}{|p{2.5cm}|}{\textbf{~~RX2~(VV), x = 30~m}} \\
		
\hline
             \multicolumn{1}{|p{.7cm}|}{\textbf{~~~~~~~~~~Param.}}&\multicolumn{1}{|p{.7cm}|}{{h=10~m}}&\multicolumn{1}{|p{.7cm}|}{{h=20~m}}&\multicolumn{1}{|p{.7cm}|}{{h=30~m}}&\multicolumn{1}{|p{.7cm}|}{{h=10~m}}&\multicolumn{1}{|p{.7cm}|}{{h=20~m}}&\multicolumn{1}{|p{.7cm}|}{{h=30~m}}&\multicolumn{1}{|p{.7cm}|}{{h=10~m}}&\multicolumn{1}{|p{.7cm}|}{{h=20~m}}&\multicolumn{1}{|p{.7cm}|}{{h=30~m}}&\multicolumn{1}{|p{.7cm}|}{{h=10~m}}&\multicolumn{1}{|p{.7cm}|}{{h=20~m}}&\multicolumn{1}{|p{.7cm}|}{{h=30~m}}\\
\hline

\multicolumn{1}{|c|}{$|\Gamma_1(\Psi)|$}&\multicolumn{1}{|c|}{{-}}&\multicolumn{1}{|c|}{{-}}&\multicolumn{1}{|c|}{{-}}&\multicolumn{1}{|c|}{{-}}&\multicolumn{1}{|c|}{{-}}&\multicolumn{1}{|c|}{{-}}&\multicolumn{1}{|c|}{{0.59}}&\multicolumn{1}{|c|}{{0.67}}&\multicolumn{1}{|c|}{{0.7}}&\multicolumn{1}{|c|}{{0.39}}&\multicolumn{1}{|c|}{{0.57}}&\multicolumn{1}{|c|}{{0.64}}\\
\hline

\multicolumn{1}{|c|}{\thead{$|c_{\rm pol}|$ [dB]\\``(UAV hovering)"}}&\multicolumn{1}{|c|}{{12.9}}&\multicolumn{1}{|c|}{{7.3}}&\multicolumn{1}{|c|}{{5.6}}&\multicolumn{1}{|c|}{{8}}&\multicolumn{1}{|c|}{{6.3}}&\multicolumn{1}{|c|}{{6.0}}&\multicolumn{1}{|c|}{{11.2}}&\multicolumn{1}{|c|}{{8.2}}&\multicolumn{1}{|c|}{{5.7}}&\multicolumn{1}{|c|}{{11.3}}&\multicolumn{1}{|c|}{{8.8}}&\multicolumn{1}{|c|}{{8.0}}\\
\hline

\multicolumn{1}{|c|}{\thead{$|c_{\rm pol}|$ [dB]\\``(UAV moving)"}}&\multicolumn{1}{|c|}{{0.6}}&\multicolumn{1}{|c|}{{0.4}}&\multicolumn{1}{|c|}{{2}}&\multicolumn{1}{|c|}{{5.4}}&\multicolumn{1}{|c|}{{4.6}}&\multicolumn{1}{|c|}{{4.3}}&\multicolumn{1}{|c|}{{1.8}}&\multicolumn{1}{|c|}{{0.7}}&\multicolumn{1}{|c|}{{2.2}}&\multicolumn{1}{|c|}{{5.8}}&\multicolumn{1}{|c|}{{2.5}}&\multicolumn{1}{|c|}{{4.1}}\\
\hline
		\end{tabular}
\caption{Ground reflection coefficient, $|\Gamma_1(\Psi)|$, and polarization mismatch loss factor ratio, $|c_{\rm pol}|$, values obtained from measurement data. The value of $|c_{\rm pol}|$, for unobstructed UAV hovering scenario and UAV moving scenario.} \label{Table:Polarization_loss}
\end{table*}

\begin{figure}[!t]
	\centering
	\includegraphics[width=0.6\columnwidth]{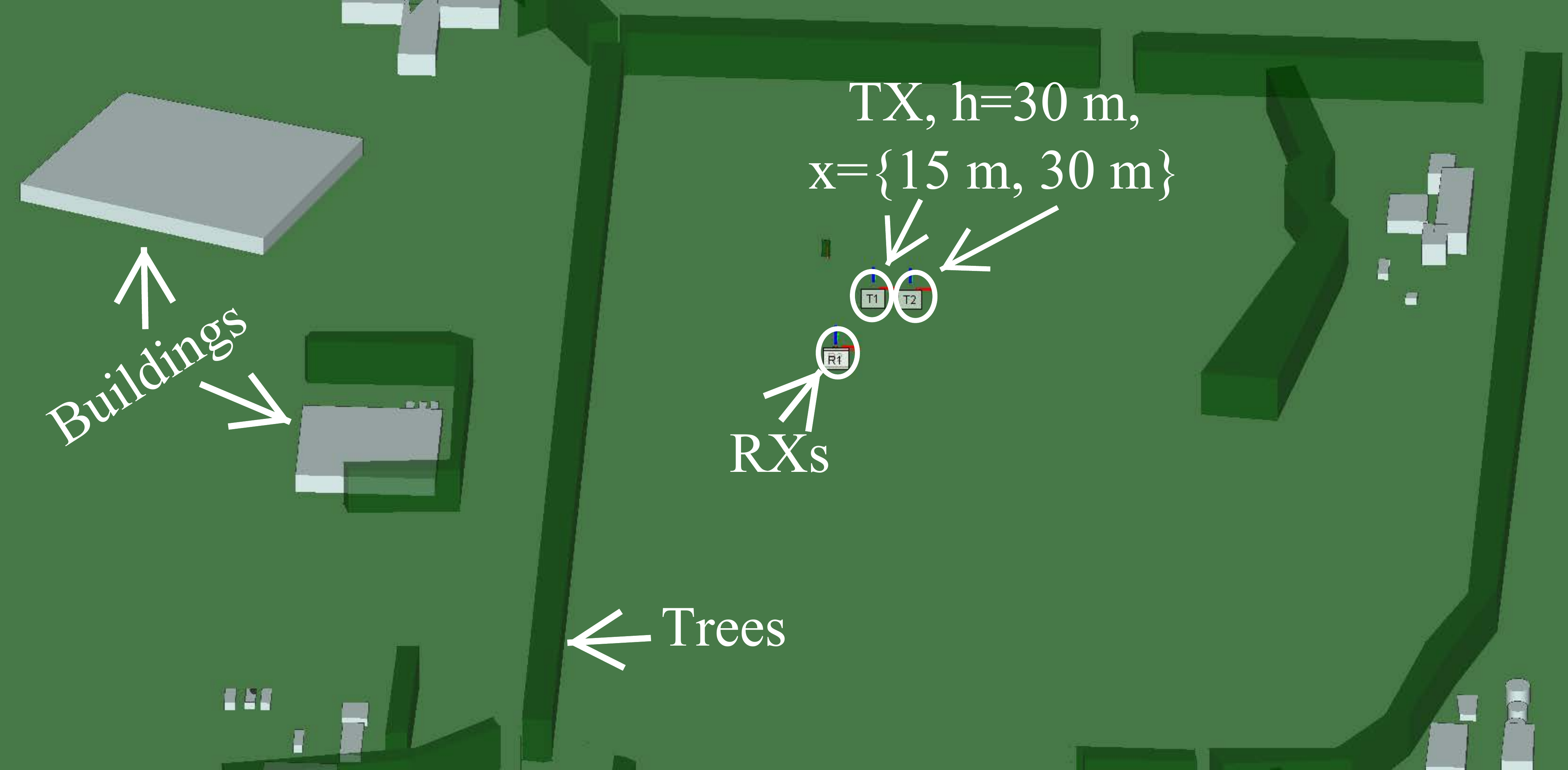}
	\caption{The simulation setup of the measurements created using Wireless InSite~(ray tracing software).}\label{Fig:RT_scenario}
\end{figure}

\subsection{Ray Tracing Simulation Setup and Path Loss Results}

Ray tracing simulations were carried out using the Wireless InSite software to compare with our empirical findings for unobstructed UAV hovering scenario. The foliage obstructed scenario was not simulated using ray tracing due to the limitation of creating a particular real-world tree structure in simulations for UWB propagation. Similarly, the unobstructed UAV moving scenario was also not simulated, as the effect of the motion on a given UWB propagation environment cannot be easily captured. 

\begin{figure}[!t]
		\centering
	\includegraphics[width=\textwidth]{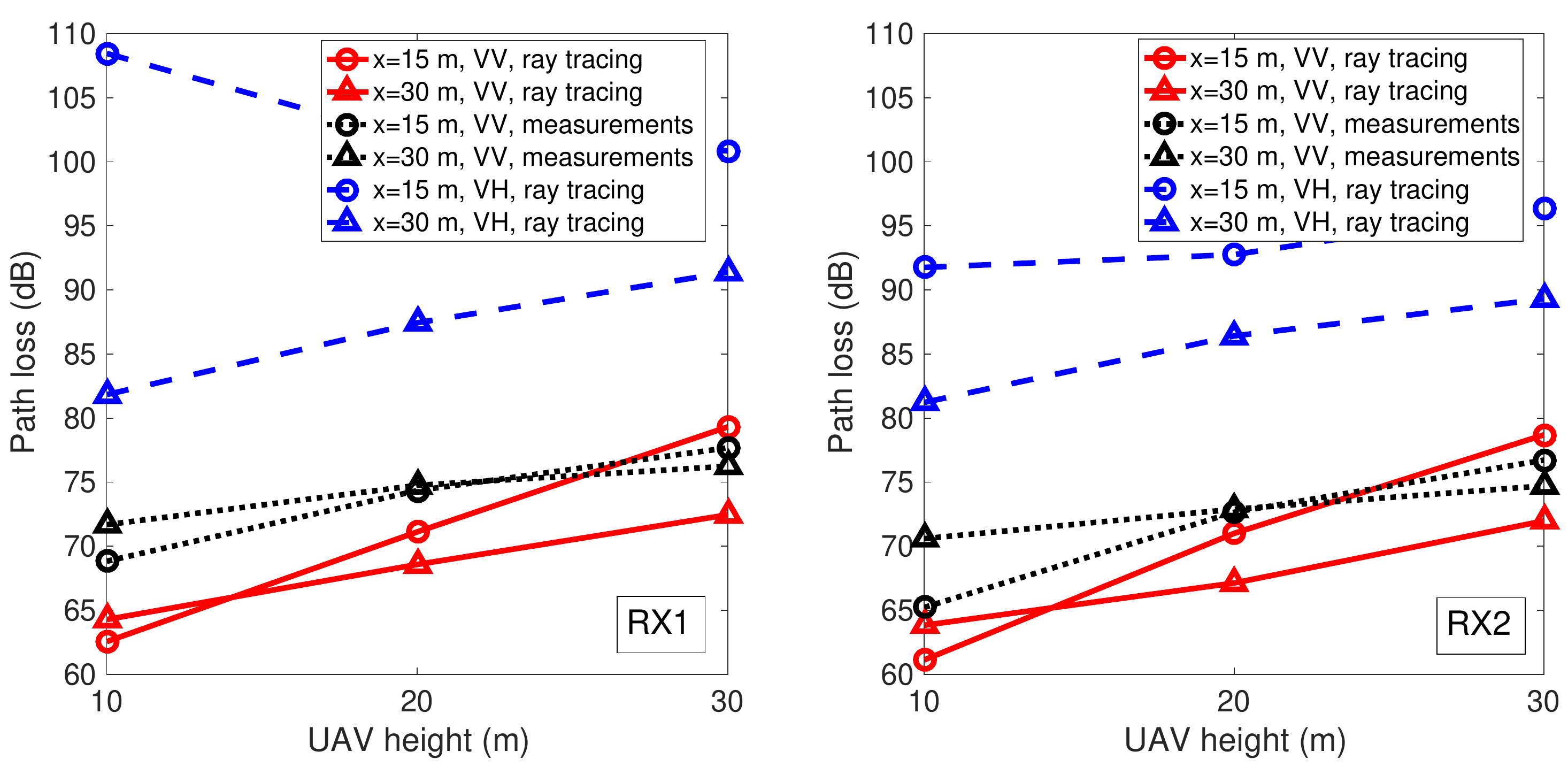} 
	\caption{(a) Ray tracing simulation results of path loss (VV and VH) alongside measurement results (VV) for unobstructed UAV hovering at $x=15$~m, and $x=30$~m for different heights for (a) RX1, at ground, (b) RX2, at $1.5$~m height from the ground.}\label{Fig:attenuation_meas_raytracing}
\end{figure}

Fig.~\ref{Fig:RT_scenario} shows the simulation environment created using ray tracing. The majority of the MPCs arise from scatterers near the RXs, e.g., tripod, measuring equipment, and humans that were modeled accordingly. The MPCs from the buildings and foliage in the surroundings were not considered as our empirical excess time delay for each scan is limited to $100$~ns~($30$~m). This corresponds to reflections only from the objects that are not farther than $15$~m~(considering two-way reflection distance $2\times15$~m) from the RXs. Similarly, due to the limited time resolution of MPCs in the ray tracing software compared to measurements, only path loss results are considered for comparison. 

The path loss results obtained from ray tracing simulations for unobstructed UAV hovering scenario for VV and VH antenna orientations are shown in Fig.~\ref{Fig:attenuation_meas_raytracing}. Measurement results of unobstructed UAV hovering for VV antenna orientation are also provided in Fig.~\ref{Fig:attenuation_meas_raytracing}. Comparing with the empirical results at RX1 and RX2, respectively, we observe a reasonably close match. Cross overs between the path loss results at $15$~m and $30$~m for VV scenario can also be observed in ray tracing simulations, similar to measurement and analytical results. However, we observe smaller path loss at smaller UAV heights compared to empirical results. Similarly, larger path loss can be observed at larger elevation angles compared to empirical results. This is mainly due to the perfect donut shaped antenna radiation pattern in simulations.

For the VH antenna orientation, we found that the path loss from ray tracing simulations  differs from that observed empirically in Fig.~\ref{Fig:PL_mean_VH}(a). This is because the MPCs~(with weak or no LOS component) obtained due to reflections from surrounding objects are dependent on the actual geometry and placement of these objects, which we have not modeled with high precision. In addition, the path loss obtained for VH antenna orientation is significantly larger than measurements. First, the additional scatterers not modeled in ray tracing but present in the actual measurements contribute to additional MPCs. This results in a reduction of path loss. Another reason is the non-ideal cross-polarization discrimination for real-world antennas that results in the reduction of the path loss for VH antenna orientation.

\section{Conclusion}\label{Section:Conclusions} 
In this work, we have reported UWB AG propagation channel measurements in an open field using a small UAV in three propagation scenarios: unobstructed UAV hovering, foliage obstruction, and unobstructed UAV moving in a constant-altitude circle. Measurements were obtained at three UAV heights and two horizontal distances for two different antenna orientations at the UAV TX. We observed that the received power is highly dependent on the antenna gain of the LOS component in the elevation plane when the antennas are aligned~(same orientation). The antenna gain for the LOS component can be approximated by a sine function of the elevation angle. This antenna gain was used in an analytical model for the path loss for unobstructed UAV hovering and moving scenarios. In addition, empirical path loss results for unobstructed UAV hovering scenario are corroborated using ray tracing simulations. Moreover, it was found that the antenna orientation mismatch results in higher path loss and a larger number of MPCs, smaller $K$-factor, and larger RMS-DS than in the co-polarized case. The OLOS scenario due to foliage between the TX and the RX, while the UAV is hovering, introduces additional attenuation, and additional MPCs due to foliage, resulting in further reduction in the $K$-factor. The motion of the UAV in a circular path provided better mitigation against antenna orientation mismatch than the unobstructed UAV hovering scenario. A statistical channel model based upon the SV model was derived from the empirical results. The SV model was found to provide a better fit for the PDP than a single exponential fit. Future work includes channel characterization for suburban and urban areas and air-to-air propagation channel measurements and modeling.


\bibliographystyle{IEEEtran}


\end{document}